\documentclass[aps,amsmath,amssymb,prc,twosides,twocolumn,floats,superscriptaddress]{revtex4-2}

\usepackage{graphicx} % Required for inserting images
\usepackage[utf8]{inputenc}
\usepackage[english]{babel}
\usepackage[shortlabels]{enumitem}
\usepackage{bm}
\usepackage{float}
\usepackage{tabularx}
\usepackage{array}
\usepackage[dvipsnames]{xcolor}
\usepackage{hyperref}

\usepackage{multirow, tikz}
\usepackage{longtable}
\usepackage{dcolumn}
\newcolumntype{d}{D{.}{.}{-1}}
\usepackage{setspace}
\usepackage{threeparttable}
\usepackage{tabularx}
\usepackage[symbol*]{footmisc}
\DefineFNsymbolsTM{myfnsymbols}{% def. from footmisc.sty "bringhurst" symbols
  \textasteriskcentered *
  \textdagger    \dagger
  \textdaggerdbl \ddagger
  \textsection   \mathsection
  \textbardbl    \|%
  \textparagraph \mathparagraph
}%
\setfnsymbol{myfnsymbols}

\usepackage{siunitx}
\newcommand{\nuc}[2]{\hbox{$^{#1}$#2}}

\usepackage{gensymb}
\usepackage{mathtools}

\usepackage{booktabs}
\usepackage{afterpage}
\begin{document}

\title{Experimental study of \nuc{53}{Cr} via the $(d,p\gamma)$ reaction}

\author{M. Spieker}
\email{Corresponding author: mspieker@fsu.edu}
\affiliation{Department of Physics, Florida State University, Tallahassee, Florida 32306, USA}

\author{L.A. Riley}
\affiliation{Department of Physics and Astronomy, Ursinus College, Collegeville, PA 19426}

\author{M. Heinze}
\affiliation{Department of Physics and Astronomy, Ursinus College, Collegeville, PA 19426}

\author{A.L. Conley}
\affiliation{Department of Physics, Florida State University, Tallahassee, Florida 32306, USA}

\author{B. Kelly}
\affiliation{Department of Physics, Florida State University, Tallahassee, Florida 32306, USA}

\author{P.D. Cottle}
\affiliation{Department of Physics, Florida State University, Tallahassee, Florida 32306, USA}

\author{R. Aggarwal}
\affiliation{Department of Physics, Florida State University, Tallahassee, Florida 32306, USA}

\author{S. Ajayi}
\affiliation{Department of Physics, Florida State University, Tallahassee, Florida 32306, USA}

\author{L.T. Baby}
\affiliation{Department of Physics, Florida State University, Tallahassee, Florida 32306, USA}

\author{S. Baker}
\affiliation{Department of Physics, Florida State University, Tallahassee, Florida 32306, USA}

\author{I. Conroy}
\affiliation{Department of Physics and Astronomy, Ursinus College, Collegeville, PA 19426}

\author{I.B. D'Amato}
\affiliation{Department of Physics, Davidson College, Davidson, NC 28035, USA}

\author{J. Esparza}
\affiliation{Department of Physics, Florida State University, Tallahassee, Florida 32306, USA}

\author{S. Genty}
\affiliation{Department of Physics, Florida State University, Tallahassee, Florida 32306, USA}

\author{I. Hay}
\affiliation{Department of Physics, Florida State University, Tallahassee, Florida 32306, USA}

\author{K.W. Kemper}
\affiliation{Department of Physics, Florida State University, Tallahassee, Florida 32306, USA}

\author{M.I. Khawaja}
\affiliation{Department of Physics, Florida State University, Tallahassee, Florida 32306, USA}

\author{P.S. Kielb}
\affiliation{Department of Physics, Davidson College, Davidson, NC 28035, USA}

\author{A.N. Kuchera}
\affiliation{Department of Physics, Davidson College, Davidson, NC 28035, USA}

\author{E. Lopez-Saavedra}
\altaffiliation[Current affiliation: ]{Physics Division, Argonne National Laboratory, Lemont, Illinois 60439, USA}
\affiliation{Department of Physics, Florida State University, Tallahassee, Florida 32306, USA}

\author{A.B. Morelock}
\altaffiliation[Current affiliation: ]{Department of Physics and Astronomy, University of Tennessee, Knoxville, Tennessee 37996, USA}
\affiliation{Department of Physics, Florida State University, Tallahassee, Florida 32306, USA}

\author{J. Piekarewicz}
\affiliation{Department of Physics, Florida State University, Tallahassee, Florida 32306, USA}

\author{A. Sandrik}
\affiliation{Department of Physics, Florida State University, Tallahassee, Florida 32306, USA}

\author{V. Sitaraman}
\affiliation{Department of Physics, Florida State University, Tallahassee, Florida 32306, USA}

\author{E. Temanson}
\affiliation{Department of Physics, Florida State University, Tallahassee, Florida 32306, USA}

\author{C. Wibisono}
\affiliation{Department of Physics, Florida State University, Tallahassee, Florida 32306, USA}

\author{I. Wiedenhoever}
\affiliation{Department of Physics, Florida State University, Tallahassee, Florida 32306, USA}

\date{\today}

\begin{abstract}

Excited states in \nuc{53}{Cr} were studied via the $\nuc{52}{Cr}(d,p\gamma)$ reaction up to the neutron-separation threshold. Proton-$\gamma$ angular correlations and $\gamma$ decay branching ratios were measured in particle-$\gamma$ coincidences between the Super-Enge Split-Pole Spectrograph (SE-SPS) and CeBr$_3$ Array (CeBrA) demonstrator of the John D. Fox Accelerator Laboratory at Florida State University. Previous spin-parity assignments from a $(d,p)$ singles experiment at the SE-SPS are supported and $\gamma$-ray transitions in \nuc{53}{Cr} reported. We firmly assign higher-lying excited states to \nuc{53}{Cr} because overlapping excited states and contaminants could be identified better due to the complementary $\gamma$-decay information. We also correct some of the previously reported excitation energies and present a reanalysis of previously measured $\nuc{52}{Cr}(d,p)\nuc{53}{Cr}$ angular distributions guided by the complementary $\gamma$-ray information. Based on this reanalysis, the fragmentation of the neutron $2p_{3/2}$, $2p_{1/2}$, $1f_{5/2}$, $1g_{9/2}$, and $2d_{5/2}$ single-particle strengths is reassessed for \nuc{53}{Cr}. A comparison to the corresponding strengths in \nuc{55}{Fe} is presented.

\end{abstract}


\maketitle

\section{Introduction}

The stable odd-$A$, $N=29$ isotones \cite{Ril21a, Ril22a, Ril23a} and the even-$A$, $N=29$ isotone \nuc{52}{V} \cite{Hay24a} were previously studied at the Super-Enge Split-Pole Spectrograph (SE-SPS) of the John D. Fox Accelerator Laboratory \cite{Spi24a} at Florida State University (FSU) via one-neutron $(d,p)$ transfer to map out the single-particle strengths in the $fp$ shell up to the neutron-separation energy, $S_n$, of the respective nuclei. In agreement with the expected reduction of the spectroscopic strengths relative to the single-particle estimate\,\cite{Kay13a, Kay22a}, around 55(10)\,$\%$ of the single-particle strengths were collected for the $2p_{3/2}$, $2p_{1/2}$, and $1f_{5/2}$ orbitals up to $S_n$\,\cite{Spi24a}. In addition, the new data showed that the energy difference between the $2p_{1/2}$ and $1f_{5/2}$ indeed decreases with increasing proton number $Z$, possibly explaining the vanishing of the $N=32$ subshell gap when going from the exotic and apparently doubly magic \nuc{52}{Ca} to stable \nuc{58}{Fe} (see, {\it e.g.}, \cite{Pri01a, Lei18a, Liu19a}). As discussed in \cite{Spi24a}, it is the centroid energy of the $2p_{1/2}$ orbital though which increases rather than the $1f_{5/2}$ significantly coming down in energy. In fact, the data supported the conclusion that the $1f_{5/2}$ centroid energy remained almost constant in \nuc{51}{Ti}, \nuc{53}{Cr}, and \nuc{55}{Fe} (see compilation in \cite{Spi24a}). 

\begin{figure*}[t]
    \centering
    \includegraphics[width=1.0\linewidth]{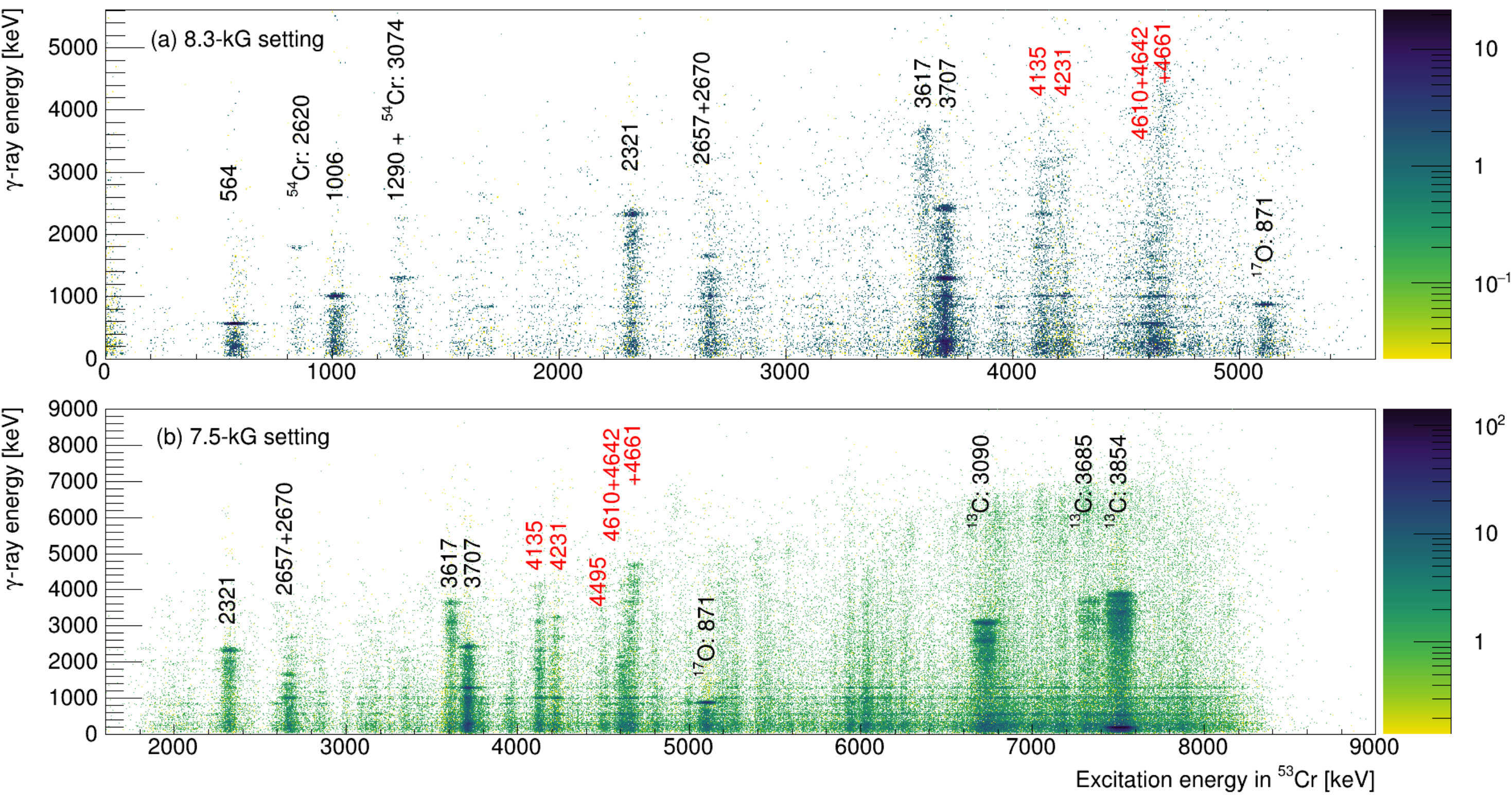}
    \caption{Proton-$\gamma$ coincidence matrix for (a) 8.3-kG and (b) 7.5-kG magnetic settings of the SE-SPS in $\nuc{52}{Cr}(d,p\gamma)\nuc{53}{Cr}$. Some excited states of \nuc{53}{Cr} are highlighted with their excitation energy given in keV. As a natural chromium target on a carbon backing was used for the experiment, excited states of \nuc{54}{Cr}, \nuc{17}{O} and \nuc{13}{C} can also be identified in the coincidence matrix. For states marked in red, excitation energies reported in \cite{Ril23a} are corrected. See text for further discussion.}
    \label{fig:matrix}
\end{figure*}

Questions remained about the detected $1g_{9/2}$ spectroscopic strength. Only about 30\,$\%$ of the expected sum-rule strength were detected up to $S_n$, i.e., significantly less than for the orbitals of the $fp$ shell\,\cite{Ril23a, Hay24a}. While it was speculated that the bulk of the $1g_{9/2}$ strength could be located above $S_n$, there was a possibility that smaller fragments were missed in the dense $(d,p)$ spectra at higher energies. Our previous work on $\nuc{54}{Fe}(d,p)\nuc{55}{Fe}$\,\cite{Ril22a} and $\nuc{61}{Ni}(d,p)\nuc{62}{Ni}$\,\cite{Spi23a} also hinted at the possibility that a significant fraction of $\ell = 2$ strength associated with transferring neutrons into the $2d_{5/2}$ orbital was missed in $\nuc{52}{Cr}(d,p)\nuc{53}{Cr}$\,\cite{Ril23a}. Data in \nuc{55}{Fe} and \nuc{62}{Ni} suggest that the $\ell =2$ strength is largely fragmented. Hay {\it et al.} proposed that the coincident detection of $\gamma$ rays and associated decay-channel selection in a $(d,p\gamma)$ experiment with a high-resolution spectrograph might provide the selectivity needed to identify weaker fragments of strongly fragmented single-particle strength\,\cite{Hay24a} and could, thus, inform $(d,p)$ singles experiments.

In this work, we report new data from such a $\nuc{52}{Cr}(d,p\gamma)\nuc{53}{Cr}$ experiment performed at the John D. Fox Laboratory with the SE-SPS and CeBr$_3$ Array (CeBrA) demonstrator for coincident $\gamma$-ray detection \cite{Con24a}. The new data support spin-parity assignments, $J^{\pi}$, from previous $(d,p)$ singles experiments \cite{Boc65a, Rao68a, Ril23a} based on measured particle-$\gamma$ angular correlations and also support previous reports of $\gamma$-ray transitions assigned to $\nuc{53}{Cr}$ in $(d,p\gamma)$ \cite{Rao68a, Car70a}. We add a number of previously unobserved $\gamma$-ray transitions for higher-lying states and present multipole-mixing ratios for several transitions observed in $(d,p\gamma)$. We also use the complementary $\gamma$-ray information to clearly assign higher-lying states to $\nuc{53}{Cr}$ reassessing the large number of excited states previously assigned to $\nuc{53}{Cr}$ based on information from $(d,p)$ experiments (see, {\it e.g.}, \cite{Boc65a, Rao68a}). In addition, excitation energies higher than 4.1\,MeV, for states reported in \cite{Ril23a}, are revisited and a reanalysis of the previously measured $\nuc{52}{Cr}(d,p)\nuc{53}{Cr}$ angular distributions guided by the complementary $\gamma$-ray information is presented. The latter enabled a more precise assessment of the fragmentation of the $2p_{3/2}$, $2p_{1/2}$, $1f_{5/2}$, $1g_{9/2}$, and $2d_{5/2}$ single-particle strengths in \nuc{53}{Cr}, and a comparison to the $N=29$ isotone \nuc{55}{Fe}\,\cite{Ril22a}. Some aspects are discussed with respect to the predictions made by covariant density functional theory (CDFT) calculations employing the relativistic functionals FSUGarnet, RMF022, and RMF028\,\cite{Che15a}. We also comment on how the spin-orbit splitting between the $2p_{3/2}-2p_{1/2}$ spin-orbit partners fits into the systematics, which were presented in \cite{Che24a}. Differences in the fragmentation of the $2d_{5/2}$ neutron single-particle strength in the $N=29$ isotones \nuc{53}{Cr} and \nuc{55}{Fe} are highlighted. The data and results of the present analysis supersede the results and conclusions presented in \cite{Ril23a}.

\section{Experimental Details and Data Analysis}

\begin{figure*}[t]
    \centering
    \includegraphics[width=1.0\linewidth]{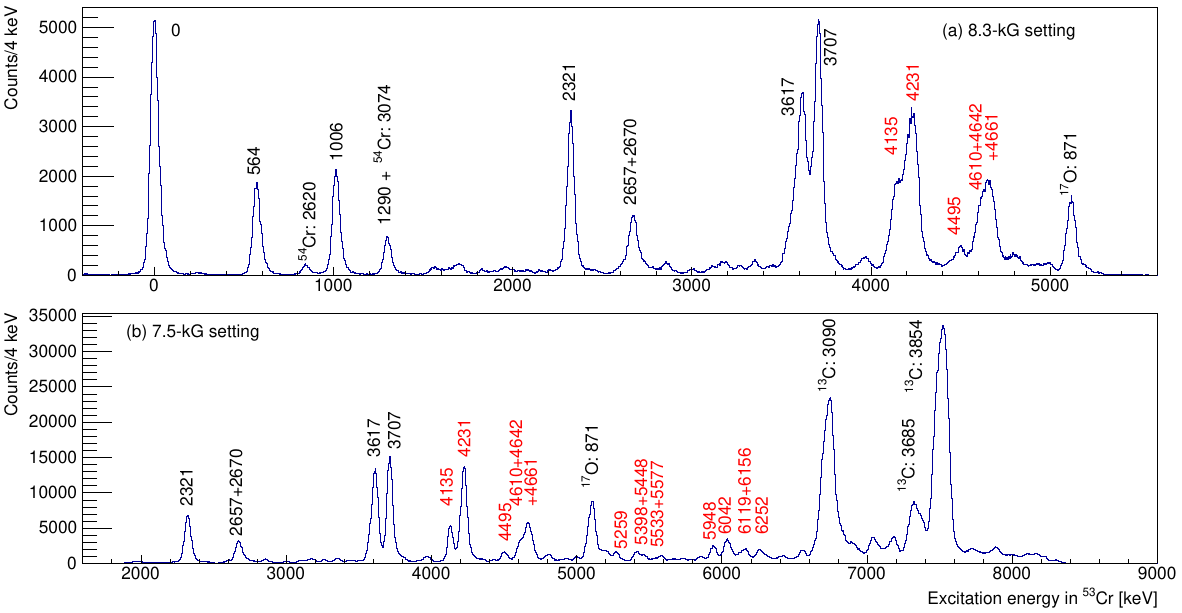}
    \caption{Proton spectra measured for (a) 8.3-kG and (b) 7.5-kG magnetic settings of the SE-SPS in $\nuc{52}{Cr}(d,p)\nuc{53}{Cr}$ at $37^{\circ}$. The excitation energy in \nuc{53}{Cr} is given on the $x$-axis. States in \nuc{53}{Cr} and reported in \cite{Ril23a} are marked with their excitation energy. For states marked in red, excitation energies reported in \cite{Ril23a} are corrected. Some contaminants are marked, too. Note that unlike in the $25^{\circ}$ spectrum shown in \cite{Ril23a} the \nuc{13}{C} and \nuc{17}{O} ground states are part of the proton groups around 3.7\,MeV and 4.2\,MeV, respectively, and not separately resolved. No $\gamma$ coincidences were required. See text for further discussion.}
    \label{fig:xavg}
\end{figure*}

The $\nuc{52}{Cr}(d,p\gamma)\nuc{53}{Cr}$ experiment was performed at the John D. Fox Superconducting Linear Accelerator Laboratory of Florida State University \cite{Spi24a}. The Fox Laboratory operates a 9-MV Super-FN Tandem van-de-Graaff accelerator. Deuterons were injected from a NEC SNICS-II cesium sputter ion source into the Tandem and accelerated up to an energy of 16\,MeV. The beam was guided towards the scattering chamber for particle-$\gamma$ coincidence experiments at the Super-Enge Split-Pole Spectrograph (SE-SPS) \cite{Spi24a}, where it impinged onto a natural chromium foil of areal density $\rho =300$\,$\mu$g/cm$^2$ evaporated onto a natural 20-$\mu$g/cm$^2$ carbon backing. Two magnetic field settings were used, 8.3\,kG and 7.5\,kG, to identify excited states of \nuc{53}{Cr} populated in the $\nuc{52}{Cr}(d,p)$ reaction up to the neutron-separation energy, $S_n = 7939.12(14)$\,keV \cite{ensdf}, with the position-sensitive focal-plane detector of the SE-SPS. The $\gamma$ decays of these excited states were studied with the CeBr$_3$ Array (CeBrA) demonstrator \cite{Con24a}, which consisted of four $2 \times 2$ inch CeBr$_3$ detectors and one $3 \times 4$ inch CeBr$_3$ detector placed around the scattering chamber in a plane with azimuthal angle $\phi_{\gamma} = 0^{\circ}$. For the particle-$\gamma$ coincidence experiment described here, the SE-SPS was placed at a laboratory scattering angle of $\theta_{\mathrm{SE-SPS}}=37^{\circ}$. Efficiency measurements for the $\gamma$-ray detectors were performed with \nuc{56}{Co}, \nuc{60}{Co}, and \nuc{152}{Eu} standard calibration sources. Data were recorded using two CAEN V1725S digitizers with DPP-PSD firmware. Coincidences between the plastic scintillator of the SE-SPS focal-plane detector and the CeBr$_3$ $\gamma$-ray detectors were acquired within a 3-$\mu$s coincidence window. In addition to these coincidence events, SE-SPS singles were recorded. In this experiment, the SE-SPS energy resolution for protons was around 50\,keV (FWHM). The solid-angle acceptance was $\Delta \Omega = 4.6$\,msr. Events were built with custom-made software \cite{con25a} based on the ROOT data analysis framework \cite{Bru97a}. Previously measured angular distributions of protons \cite{Ril23a} were modeled using the coupled-channels program \textsc{chuck3} \cite{chuck} with global optical model parameters from \cite{Kon03a}. To calculate the deuteron optical model parameters, the approach of \cite{Wal76a} was adopted. This is referred to as the adiabatic distorted wave approximation (ADWA) in literature. Scattering amplitudes from the \textsc{chuck3} calculations were used as inputs for the calculations of proton-$\gamma$ angular correlations with the computer program \textsc{angcor} \cite{angcor}, which follows the formalism presented in \cite{Ryb70a}. Details for the combined SE-SPS+CeBrA demonstrator setup were discussed in \cite{Con24a, Spi24a}. Further information on the $\nuc{52}{Cr}(d,p)\nuc{53}{Cr}$ singles experiments, of which some aspects will be discussed in this work, can be found in \cite{Ril23a}.

%For, {\it e.g.}, \nuc{53}{Cr}, the conclusions drawn rely, however, on the assumption that all $\ell =3$ and $\ell=1$ strengths observed at excitation energies larger than $E_x = 1.5$\,MeV and 2.3\,MeV, respectively, can be associated with single-particle excitations placing the transferred neutron in either the $1f_{5/2}$ ($\ell=3$) or $2p_{1/2}$ ($\ell =1$) orbital. Even in the absence of strong primary $\gamma$-ray transitions and as we previously discussed \cite{Con24a, Spi24a}, spin-orbit partners can potentially be distinguished through secondary $\gamma$-ray transitions and associated particle-$\gamma$ angular correlations, which would provide the means to resolve possible ambiguities.

\section{Results and Discussion}

\begin{figure*}[t]
    \centering
    \includegraphics[width=1.0\linewidth]{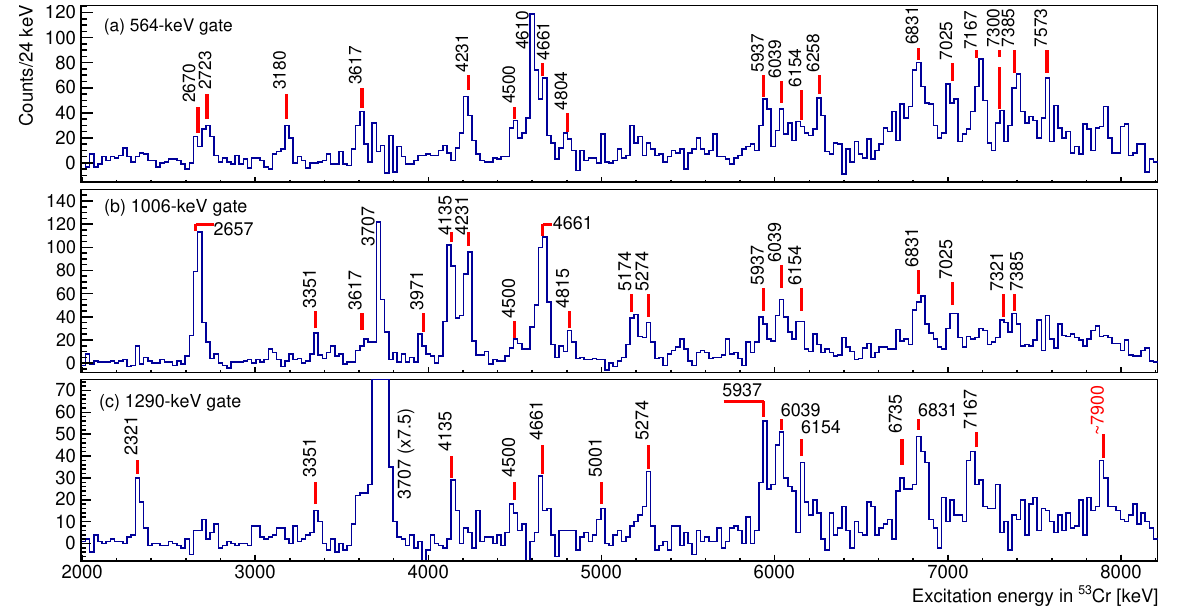}
    \caption{$\gamma$-ray gated excitation-energy (proton) spectra. Horizontal gates in the proton-$\gamma$ coincidence matrix of Fig.\,\ref{fig:matrix}\,(b) were set on the (a) 564-keV ($1/2^-_1 \rightarrow 3/2^-_1$), (b) 1006-keV ($5/2^-_1 \rightarrow 3/2^-_1$), and (c) 1290-keV ($7/2^-_1 \rightarrow 3/2^-_1$) $\gamma$-ray transitions in \nuc{53}{Cr}. Some excited states of \nuc{53}{Cr} are highlighted with their adopted excitation energy. Red markers are set at the exact energies. As can be seen, the gates clearly filter states out differently. Note that observing the excited states with these gates applied does not mean that they decay directly to the states to which the $\gamma$-ray transition of the gate belongs. Secondary transitions are possible. The peak in panel (c) at $\sim 7900$\,keV has not been observed before. It corresponds to the 7891-keV state established in this work based on the $(d,p)$ singles data, see Sec.\,\ref{sec:singles}. Also see text for more details.}
    \label{fig:proj}
\end{figure*}

The proton-$\gamma$ coincidence matrix measured in the $(d,p\gamma)$ coincidence experiment with the SE-SPS and CeBrA demonstrator is shown in Fig.\,\ref{fig:matrix}. Some excited states of \nuc{53}{Cr} are highlighted with their excitation energy given in keV. Additionally, states of $\nuc{13}{C}$ coming from $(d,p)$ reactions on the carbon backing, and states of \nuc{17}{O} and \nuc{54}{Cr} originating from $(d,p)$ reactions on target contaminants are marked. Note that \nuc{52}{Cr} constitutes 83.789\,$\%$ of natural chromium, while \nuc{53}{Cr} contributes with 9.501\,$\%$ \cite{nndc}. Reactions on other chromium isotopes could not be identified above background.

\begin{figure}[t]
    \centering
    \includegraphics[width=0.95\linewidth]{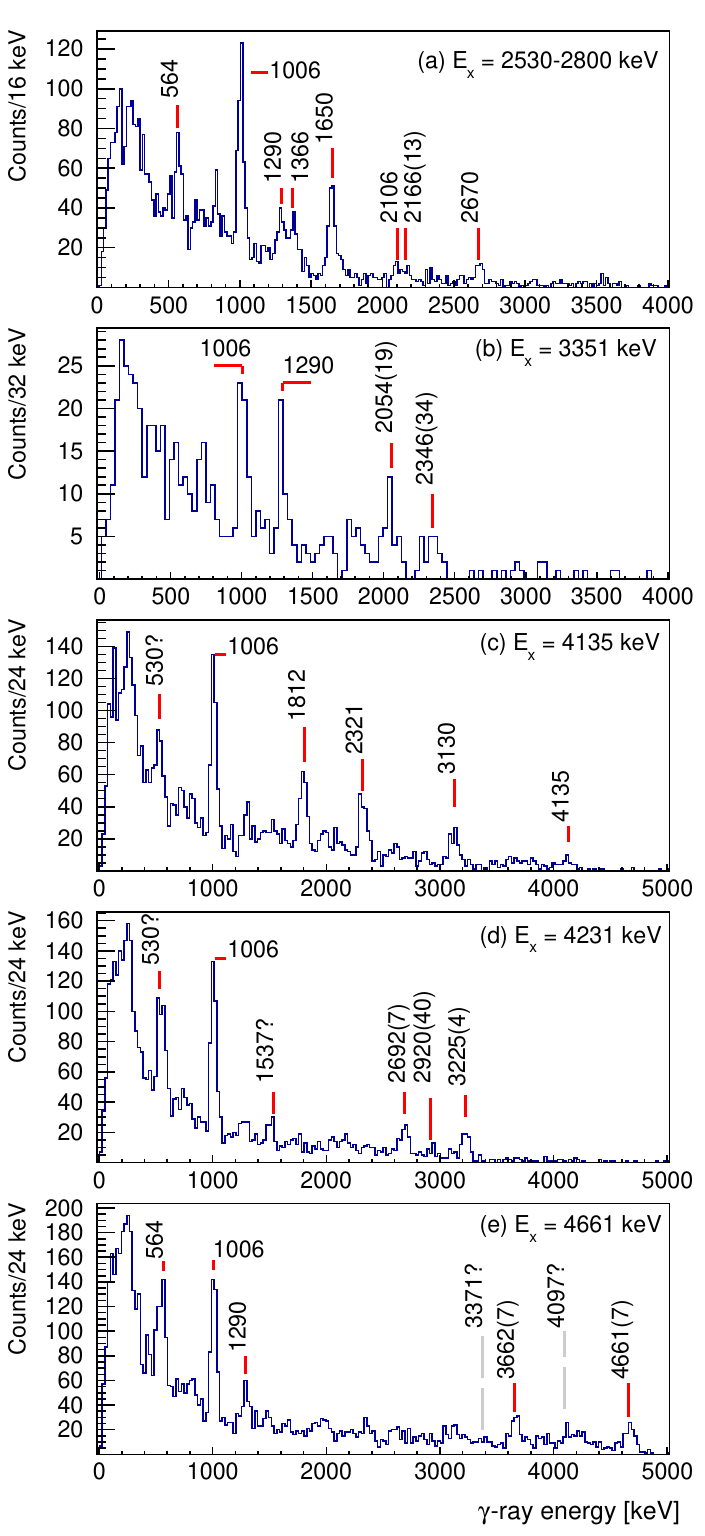}
    \caption{(a)-(e) observed $\gamma$-ray transitions when different excitation-energy gates are applied. The selected excitation energies are indicated in the panels. Uncertainties are stated for newly observed $\gamma$-ray transitions or any transition for which corrections seem necessary. See text for more details.}
    \label{fig:gammas}
\end{figure}

The complementary $\gamma$-ray information enabled a more precise determination of excitation energies in \nuc{53}{Cr}. Fig.\,\ref{fig:xavg} shows the proton spectra measured with the SE-SPS at the two different magnetic settings mentioned above. States of \nuc{53}{Cr} are highlighted with their excitation energy in keV. States for which we report a correction of their excitation energy from Ref.\,\cite{Ril23a} are additionally highlighted in red; see also Fig.\,\ref{fig:matrix}\,(b) for some of them. For states reported in \cite{Ril23a}, information on excitation energies higher than 4.1\,MeV needs to be revisited. We will come back to this discussion in section \ref{sec:singles}. To identify states uniquely as excited states of \nuc{53}{Cr}, we applied horizontal gates to the proton-$\gamma$ coincidence matrix. Specifically, gates were set on the $\gamma$-ray transitions depopulating the three lowest excited states of \nuc{53}{Cr} at 564\,keV, 1006\,keV, and 1290\,keV\,\cite{Jun09a}, i.e., states through which most of the higher-lying states will funnel should they not directly decay to the ground state. The resulting proton spectra for the 7.5-kG setting when these conditions are applied, respectively, are shown in Fig.\,\ref{fig:proj}. Background was subtracted from these spectra with gates set just above and below the mentioned $\gamma$-ray transitions. The widths of these background gates were half the width of the gates selecting the associated $\gamma$-ray transition. Several states, including even weakly populated states which cannot be clearly resolved in the proton singles spectrum of Fig.\,\ref{fig:xavg}, can be identified in Fig.\,\ref{fig:proj} and highlight the selectivity of the additional $\gamma$-ray gates. Contaminants are suppressed effectively; see, {\it e.g.}, the strongly populated states of \nuc{13}{C} in Figs.\,\ref{fig:matrix} and \ref{fig:xavg}. The agreement of the observed excitation energies with the adopted ones of \nuc{53}{Cr} \cite{Jun09a} is excellent within the resolution of both the SE-SPS and CeBrA detectors. More details will be discussed in Secs. \ref{sec:coincidences} and \ref{sec:singles}.

\subsection{Discussion of $\nuc{52}{Cr}(d,p\gamma)\nuc{53}{Cr}$ results}
\label{sec:coincidences}

\begin{figure*}[t]
    \centering
    \includegraphics[width=0.99\linewidth]{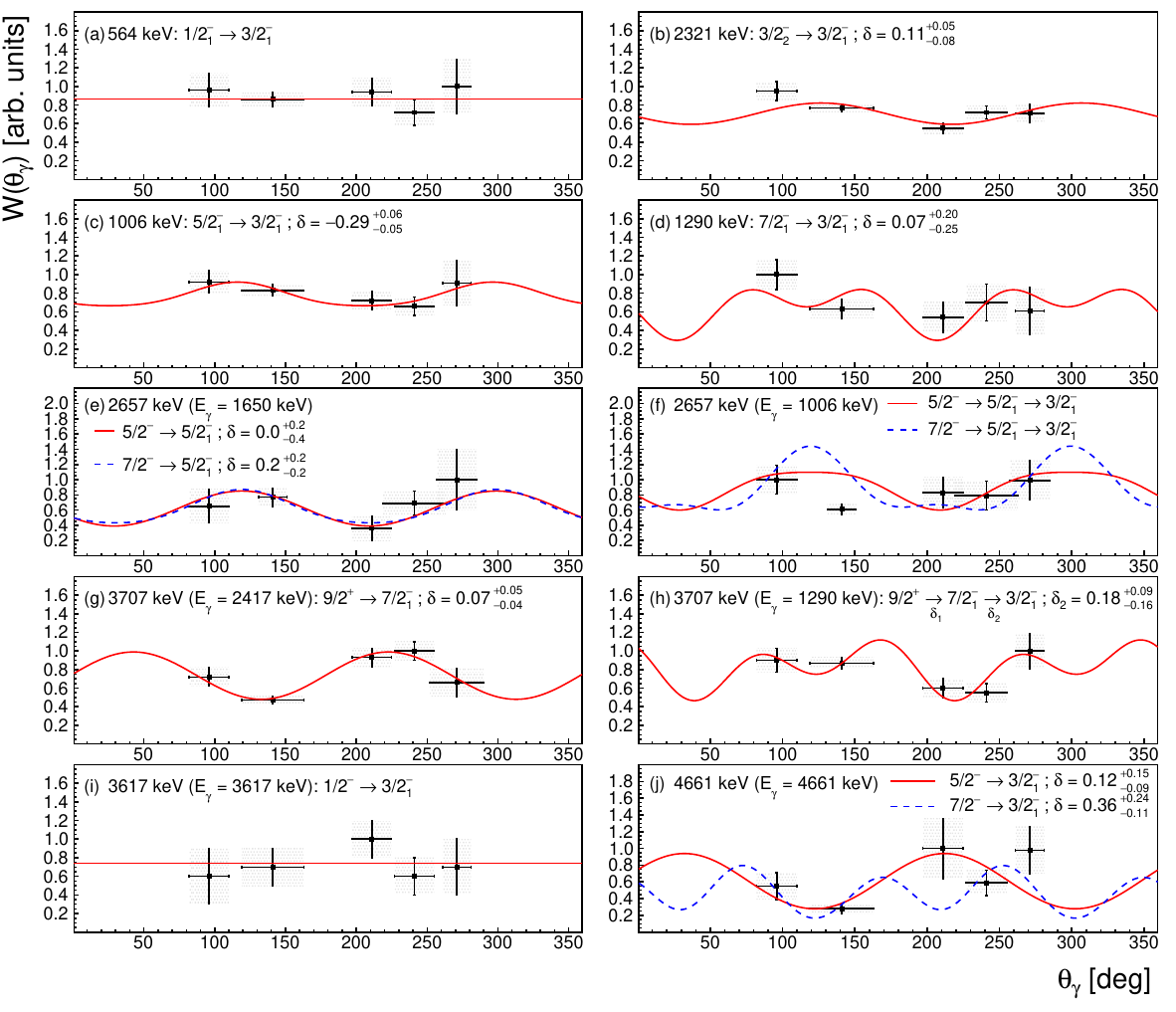}
    \caption{(a)-(j) proton-$\gamma$ angular correlations measured for excited states of \nuc{53}{Cr} in $\nuc{52}{Cr}(d,p\gamma)\nuc{53}{Cr}$ (data points with uncertainties). The excitation energies of the excited states are given in the panels. Theoretical angular correlations calculated with \textsc{angcor} are shown as lines. The multipole-mixing ratios, $\delta$, used in these calculations are also specified. All panels but (f) and (h) state the multipole-mixing ratio for the primary transition. The $\gamma$-ray energy is specified if multiple transitions are possible. In panel (h), the multipole-mixing ratio, $\delta_2$, for the secondary transition from the $7/2^-_1$ excited state to the $3/2^-_1$ ground state of \nuc{53}{Cr} is given. Panel (f) shows the secondary $5/2^-_1 \rightarrow 3/2^-_1$ transition following the decay of the 2657-keV state shown in panel (e). See text for further discussion.}
    \label{fig:angcor}
\end{figure*}

Results from the $\nuc{52}{Cr}(d,p\gamma)\nuc{53}{Cr}$ experiment will be discussed in this subsection. Some proton-$\gamma$ angular correlations were already featured in \cite{Con24a, Spi24a}. Further details are provided here including $\gamma$-decay branching ratios and multipole-mixing ratios, $\delta$, determined in our $(d,p\gamma)$ experiment. Selective, narrow excitation-energy gates were applied to study the $\gamma$ decays of excited states of \nuc{53}{Cr} and to measure particle-$\gamma$ angular correlations. Exemplary $\gamma$-ray spectra are shown in Fig.\,\ref{fig:gammas}. Proton-$\gamma$ angular correlations are presented in Fig.\,\ref{fig:angcor}. First, the $\gamma$ decay of some excited states of \nuc{53}{Cr} will be discussed. Data are summarized in table\,\ref{tab:01}.

%\subsubsection{$\gamma$-decay behavior of excited states in \nuc{53}{Cr}}

As can already be seen in Figs.\,\ref{fig:proj}\,(a) and (b), at least three states are populated around an excitation of 2.7\,MeV. They are the adopted 2657-keV, 2670-keV, and 2723-keV states. The $\gamma$ rays from the states in this region are shown in Fig.\,\ref{fig:gammas}\,(a). By calculating the efficiency-corrected yields for the 1366-keV and 1650-keV transitions, which are primary transitions depopulating the 2657-keV state, $\gamma$-decay intensities of 42(7) and 100(7) are determined, respectively. Within uncertainties, these are in excellent agreement with the adopted values of 39(10) and 100(10)\,\cite{Jun09a}. 

Two $\gamma$ rays depopulating the 2670-keV state are observed. These are the previously reported 2106-keV and 2670-keV transitions [see Fig.\,\ref{fig:proj}\,(a)]. We observe them at 2101(9)\,keV and 2674(7)\,keV. For them, $\gamma$-decay intensities of 60(8) and 100(8) are determined, respectively. They agree with the adopted values of $I_{\gamma} = 67$ and 100, for which, however, no uncertainties are currently listed\,\cite{Jun09a,ensdf}.

For the 2723-keV state, one $\gamma$ ray at 2166(13)\,keV was observed. No $\gamma$ transitions are adopted for this state yet. The observed transition leads to the 564-keV, $1/2^-_1$ state and establishes a state at 2730(13)\,keV. It is identified with the previously reported 2723-keV state and in agreement with $E_x = 2725(7)$\,keV determined from our $(d,p)$ singles data, see Sec.\,\ref{sec:singles}. The observed transition to the 564-keV state is also in line with the currently adopted spin-parity assignment $J^{\pi} = 1/2^-, 3/2^-$ for the 2723-keV state and the $\ell = 1$ angular momentum transfer observed in our $(d,p)$ experiment, see Sec.\,\ref{sec:singles}.

We note that the counts observed for the secondary 564-keV, 1006-keV, and 1290-keV transitions in Fig.\,\ref{fig:gammas}\,(a) are consistent with the counts from the primary transitions feeding the associated states. Counts of 147(49), 291(26), and 103(15) are expected. 169(30), 293(26), and 100(15) counts are observed, respectively.

A state with an excitation energy of 3351(6)\,keV is currently adopted and was observed with several probes\,\cite{Jun09a}. However, no $\gamma$-ray transitions are listed for that state. Figs.\,\ref{fig:proj}\,(b) and (c) suggest that the 3351-keV state decays to the 1006-keV, $5/2^-_1$ and 1290-keV, $7/2^-_1$ states. Indeed, when setting a narrow excitation-energy gate around the state, which is comparably weakly populated in $(d,p)$, two structures show up at the correct energies that likely correspond to the 2345-keV and 2061-keV transitions leading to the 1006-keV and 1290-keV states, respectively [see Fig.\,\ref{fig:gammas}\,(b)]. For these primary transitions, $\gamma$-ray energies of $E_{\gamma} = 2346(34)$\,keV and 2054(19)\,keV were determined, thus, establishing an excitation energy of 3348(39)\,keV in agreement with the adopted value and $E_x = 3349(5)$\,keV determined from our $(d,p)$ singles data. The 1006-keV and 1290-keV $\gamma$-ray lines, i.e., the secondary transitions are clearly observed, too, as can be seen in Fig.\,\ref{fig:gammas}\,(b). The $\gamma$-decay intensities can be deduced by either using the two primary or the secondary transitions. Averaging over the two results, which agree within uncertainties, yields 83(28) for the 2346-keV transition and 100(28) for the 2054-keV transition.

The 4135-keV and 4231-keV states are resolved in our spectra (see Figs.\,\ref{fig:matrix} and \ref{fig:xavg}). Fig.\,\ref{fig:proj} also proves that they partly decay to different low-lying states. The adopted $\gamma$-decay intensities for the 4135-keV state come from a previous $(d,p\gamma)$ experiment\,\cite{Car70a}. Three $\gamma$-ray transitions are listed with energies and associated intensities of 1812\,keV [$I_{\gamma} = 100(9)$], 3130\,keV [$I_{\gamma} = 80(7)$], and 4136\,keV [$I_{\gamma} = 37(5)$]\,\cite{Jun09a}. Transition energies are given without uncertainties. In our work, $\gamma$-ray transitions with energies of 1810(4)\,keV, 3121(6)\,keV, and 4119(12)\,keV are observed when setting a narrow excitation-energy gate around the 4135-keV state. The transitions can be clearly seen in the proton-$\gamma$ coincidence matrix shown in Fig.\,\ref{fig:matrix} and in Fig.\,\ref{fig:gammas}\,(c). The $\gamma$-ray transitions establish the state at $E_x = 4126(14)$\,keV in agreement with the adopted energy and $E_x = 4129(5)$\,keV from our $(d,p)$ singles data. The corresponding $\gamma$-decay intensities determined from our work are 100(17), 84(15), and 25(8). They agree well with the previously reported values.

\begingroup
\squeezetable
\renewcommand*{\arraystretch}{1.2}
\begin{table*}[!t]
\caption{\label{tab:01} Data table summarizing results from the $\nuc{52}{Cr}(d,p\gamma)\nuc{53}{Cr}$ experiment discussed in the text. Calculated excitation energies, $E_x$, and observed $\gamma$-ray energies, $E_{\gamma}$, are given. In addition, the observed $\gamma$-decay intensities, $I_{\gamma}$, are compared to the adopted intensities\,\cite{ensdf}. Spin-parity assignments, $J^{\pi}$, are from \cite{ensdf} unless otherwise noted. Multipole-mixing ratios, $\delta$, determined for some of the $\gamma$-ray transitions are presented in Fig.\,\ref{fig:angcor}. See the text for further discussion.}
\begin{ruledtabular}
\begin{tabular}{ccccccccc}

\multicolumn{2}{c}{$E_{x}$ [keV]} & \multicolumn{1}{c}{$J^{\pi}_i$} &\multicolumn{2}{c}{$E_{\gamma}$ [keV]} & \multicolumn{1}{c}{$E_{f}$ [keV]} & \multicolumn{1}{c}{$J^{\pi}_f$} & \multicolumn{2}{c}{$I_{\gamma}$}  \\
\cline{1-2} \cline{4-5} \cline{8-9}
\multicolumn{1}{c}{Ref.\,\cite{ensdf}} & \multicolumn{1}{c}{This work} & \multicolumn{1}{c}{Ref.\,\cite{ensdf}} & \multicolumn{1}{c}{Ref.\,\cite{ensdf}} & \multicolumn{1}{c}{This work} & \multicolumn{1}{c}{Ref.\,\cite{ensdf}} & \multicolumn{1}{c}{Ref.\,\cite{ensdf}} & \multicolumn{1}{c}{Ref.\,\cite{ensdf}} & \multicolumn{1}{c}{This work} \\
\hline
2656.5(3) & 2664(17)  & $5/2^-,7/2^-$ & 1366.4(5)  & 1377(15) & 1289.52(7)    & $7/2^-_1$ & 39(10)       & 42(7) \\
 & & & 1650.4(4) & 1655(8) & 1006.27(5) & $5/2^-_1$ & 100(10) & 100(7) \\
2669.9(5) & 2670(11) & $1/2^-$ & 2105.5 & 2101(9) & 564.03(4) & $1/2^-_1$ & 67 & 60(8) \\
 & & & 2670.8(7) & 2674(7) & 0 & $3/2^-_1$ & 100 & 100(8) \\
 2723(10) & 2730(13) & $1/2^-,3/2^-$ & $-$ & 2166(13) & 564.03(4) & $1/2^-_1$ & $-$ & 100 \\
 3351(6) & 3348(39) & $5/2^-,7/2^-$ & $-$ & 2054(19) & 1289.52(7) & $7/2^-_1$ & $-$ & 100(28) \\
  & & & $-$ & 2346(34) & 1006.27(5) & $5/2^-_1$ & $-$ & 83(28) \\
4135.1(6) & 4126(14) & $5/2^{+a}$ & 1812 & 1810(4) & 2320.7(2) & $3/2^-_2$ & 100(9) & 100(17) \\
 & & & 3130 & 3121(6) & 1006.27(5) & $5/2^-_1$ & 80(7) & 84(15) \\
 & & & 4136 & 4119(12) & 0 & $3/2^-_1$ & 37(5) & 25(8) \\
 4230.5(7) & 4223(41) & $5/2^{+a}$ & $-$ & 2692(7) & 1536.62(7) & $7/2^-_2$ & $-$ & 80(26) \\
 & & & 2943 & 2920(40) & 1289.52(7) & $7/2^-_1$ & 19(3) & 19(16) \\
 & & & 3222 & 3225(4) & 1006.27(5) & $5/2^-_1$ & 100(10) & 100(30) \\
 4661(7) & 4665(10) & $5/2^{-b}$ & $-$ & 3662(7) & 1006.27(5) & $5/2^-_1$ & $-$ & 98(13) \\
 & & & $-$ & 4661(7) & 0 & $3/2^-_1$ & $-$ & 100(13)

\end{tabular}
\end{ruledtabular}
\begin{flushleft}
    \footnotemark{{\small $(d,p)$ angular distribution follows $\ell=2$ neutron transfer. Neutron transfer to $2d_{5/2}$ orbital is most likely scenario. Neutron transfer to higher-lying $2d_{3/2}$ orbital is unlikely. Thus, $J^{\pi} = 5/2^+$ assignment is preferred.}}\\
    \footnotemark{{\small The $J^{\pi}$ assignment is supported by the observed $p\gamma$-angular correlation. See text for discussion.}}
\end{flushleft}
\end{table*}
\endgroup

As for the 4135-keV state, $\gamma$-decay information for the 4230-keV state comes from a previous $(d,p\gamma)$ experiment. Ref.\,\cite{Car70a} reported $\gamma$-ray transitions and associated intensities of 2943\,keV [$I_{\gamma} = 19(3)$] and 3222\,keV [$I_{\gamma} = 100(10)$]. No uncertainties were reported for the energies. We clearly observe the 3222-keV transition and see some excess of counts at $E_{\gamma} = 2920(40)$\,keV, which likely corresponds to the 2943-keV transition. In addition, there is a clear peak at 2692(7)\,keV, see Fig.\,\ref{fig:gammas}\,(d), which was not observed before. This transition might lead to the 1537-keV, $J^{\pi} = 7/2^-$ state. If true and since we observe $E_{\gamma} = 3225(4)$\,keV and 2920(40)\,keV, an excitation energy of 4223(41)\,keV would be determined in agreement with the adopted energy and 4220(6)\,keV deduced from our $(d,p)$ singles data. It is clear that the 2692-keV and 1537-keV lines only show up in the 4230-keV excitation-energy gate but not in the 4135-keV gate [see Figs.\,\ref{fig:gammas}\,(c) and (d)]. However, it must be noted that our data would indicate that the adopted decay intensities for the 1537-keV state are incorrect. Specifically, the ground-state decay branch appears to be more intense. We chose to not determine a value here as we do not claim to have the needed sensitivity for this state. Note that the 530-keV $\gamma$-ray line in Figs.\,\ref{fig:gammas}\,(c) and (d) would then correspond to the transition of the 1537-keV state to the 1006-keV state. As can be seen by comparison to Fig.\,\ref{fig:gammas}\,(e), the peak indeed appears slightly shifted compared to the 564-keV line. If our hypothesis is correct, then the decay intensities for the 4230-keV state are $I_{\gamma} = 80(26)$ for the 2692-keV transition, 19(16) for the 2943-keV transition, and 100(30) for the 3222-keV transition. Within their large uncertainties, the latter two are in good agreement with the adopted values.

Gating on the 4661-keV state returns two $\gamma$ rays which can be clearly assigned to it, see Fig.\,\ref{fig:gammas}\,(e). These are the $\gamma$ rays with energies of $E_{\gamma} = 3662(7)$ and 4661(7)\,keV. They correspond to decays to the 1006-keV, $5/2^-_1$ state and to the $3/2^-_1$ ground state of \nuc{53}{Cr}. For these transitions, we report $\gamma$-decay intensities of 98(13) and 100(13), respectively. The $\gamma$-ray transitions establish the state at $E_x = 4665(10)$\,keV. In Sec.\,\ref{sec:singles}, we will provide further evidence that this state corresponds to the adopted 4661-keV state based on the observation of an $\ell =3$ angular momentum transfer. However, as will be discussed in that section, a combined fit for the excited states at 4610\,keV, 4642\,keV, and 4661\,keV was necessary to describe the measured $(d,p)$ angular distribution. The measured proton-$\gamma$ angular correlation is consistent with a $J^{\pi} = 5/2^-$ assignment and proves that the observed ground-state decay comes from the 4661-keV state [see Fig.\,\ref{fig:angcor}\,(j)]. Proton-$\gamma$ angular correlations will be discussed next. 

As mentioned in \cite{Con24a, Spi24a}, the multipole-mixing ratios, $\delta$, are the only free parameters for determining the shape of the proton-$\gamma$ angular correlations $W(\theta_{\gamma})$ shown in Fig.\,\ref{fig:angcor}. For the case of the primary transitions, they were determined via standard $\chi^2$ minimization. Stated uncertainties for the multipole-mixing ratios in Fig.\,\ref{fig:angcor} correspond to the $1\sigma$ confidence interval. The experimental values for $W(\theta_{\gamma})$ correspond to the full-energy peak efficiency corrected, relative $\gamma$-ray yields measured at the different $\gamma$-ray detector positions.

Figs.\,\ref{fig:angcor}\,(a)-(e), (g), (i), and (j) show proton-$\gamma$ angular correlations for primary transitions orginating from the states with $E_x = 564$\,keV, 1006\,keV, 1290\,keV, 2321\,keV, 2657\,keV, 3617\,keV, 3707\,keV, and 4661\,keV. Figs.\,\ref{fig:angcor}\,(f) and (h) display secondary transitions for the 2657-keV and 3707-keV states following the primary transitions shown in Figs.\,\ref{fig:angcor}\,(e) and (g), respectively. As discussed in \cite{Spi24a}, the isotropic correlation for the 564-keV, $1/2^-_1 \rightarrow 3/2^-_1$ transition arises from the missing orientation of the $m$ substates with respect to the beam axis, i.e., the probability for populating the two projections $m_j = \pm 1/2$ is the same in $(d,p)$. This result is consistent with the flat correlations presented in \cite{Eyr74a}, where the $\gamma$-detection angle was kept fixed and the proton-detection angle varied. For the other primary transitions and for states with $J \neq 1/2$, similar to what was discussed in \cite{Spi24a}, the different population of the magnetic substates in the $(d,p)$ reaction leads to distinct proton-$\gamma$ angular correlations. For the primary transitions from the 1006-keV, 2321-keV, and 2657-keV states, we observe excellent agreement with the absolute values of the currently adopted multipole-mixing ratios\,\cite{Jun09a}. The signs are, however, opposite [see Figs.\,\ref{fig:angcor}\,(b), (c), and (e)]. We follow the sign convention of \cite{Ryb70a, angcor}.

\begin{figure*}[t]
    \centering
    \includegraphics[width=0.99\linewidth]{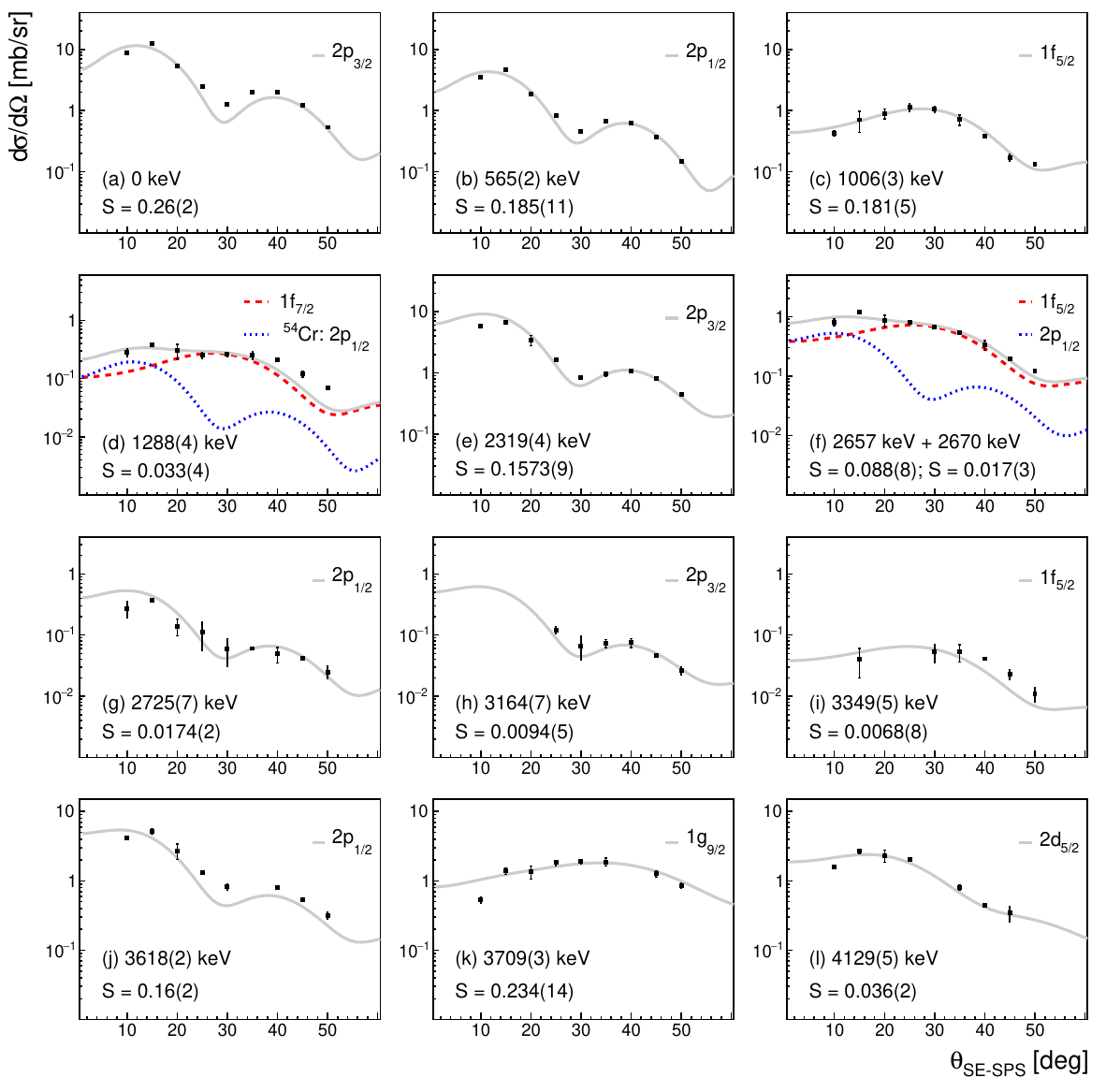}
    \caption{Angular distributions measured for excited states of \nuc{53}{Cr} in $\nuc{52}{Cr}(d,p)\nuc{53}{Cr}$. Excitation energies are given. Lines correspond to angular distributions calculated with \textsc{chuck3} and using the adiabatic distorted wave approximation. The transfer configurations used in these calculations are specified. The corresponding spectroscopic factors are also given. If two are listed, then they are given from lowest $\ell$ to highest $\ell$ transfer used in the fit. Note that this approach was adopted for doublets or triplets that could not be resolved. If two or more configurations were used for the fit, the superposition is shown as a gray line. See text for more details.}
    \label{fig:angdist_01}
\end{figure*}

To our knowledge, no value for the multipole-mixing ratio of the 1290-keV, $7/2^-_1 \rightarrow 3/2^-_1$ transition was previously reported. Considering the lowest-order multipoles possible, $E2$ and $M3$ transitions are expected to dominate. Within uncertainties, we obtain consistent values for dominant $E2$ character using a direct excitation energy gate [Fig.\,\ref{fig:angcor}\,(d)] and by studying this transition as a secondary transition following the 2417-keV, $9/2^+ \rightarrow 7/2^-_1$ transition from the 3707-keV state [Fig.\,\ref{fig:angcor}\,(h)]. For the latter, we kept $\delta_1$ fixed at the value determined for the primary 2417-keV transition [see Fig.\,\ref{fig:angcor}\,(g)]. The evaluated nuclear structure data files list an $E2$ character for this transition without a specific value for $\delta$\,\cite{ensdf}.

\begin{figure*}[t]
    \centering
    \includegraphics[width=0.99\linewidth]{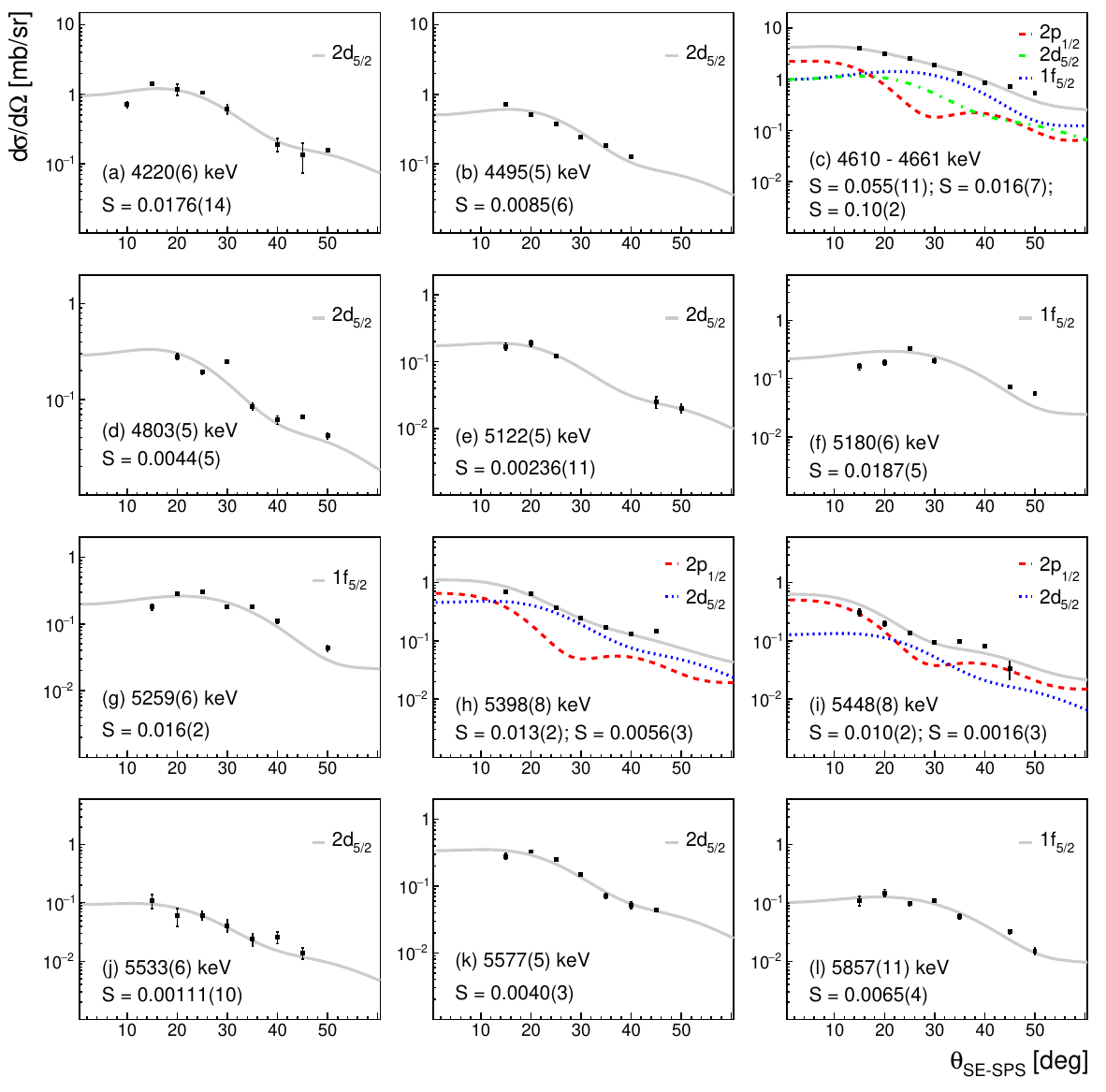}
    \caption{See caption of Fig.\,\ref{fig:angdist_01}.}
    \label{fig:angdist_02}
\end{figure*}

As can be seen in Figs.\,\ref{fig:proj}\,(a) and (b), as well as Fig.\,\ref{fig:gammas}\,(a), the 2657-keV state lies in a region where several states are populated in $(d,p)$. The statistics are sufficient to study angular correlations for the 1650-keV primary transition [Fig.\,\ref{fig:angcor}\,(e)] and 1006-keV secondary transition [Fig.\,\ref{fig:angcor}\,(f)]. For the 1650-keV transition and with the CeBr$_3$ detectors in a plane with $\phi_{\gamma}=0^{\circ}$, we cannot discriminate between a $J^{\pi} = 5/2^-$ or $7/2^-$ assignment for the 2657-keV state. Both scenarios lead to a multipole-mixing ratio of $\delta \approx 0$, indicating dominant $M1$ character for the transition in agreement with the adopted value \cite{Jun09a} from $(\alpha,n\gamma)$\,\cite{Car72a}. The value of $\delta \approx 0$ appears consistent with the correlation measured for the 1006-keV secondary transition with the exception of the data point at $\theta_{\gamma} = 141(22)^{\circ}$. While this data point might slightly favor a $J^{\pi} = 7/2^-$ assignment, the deviation from both theoretical curves prevents any firm assignment. Note that the multipole-mixing ratio for the 1006-keV secondary transition was kept fixed at the value shown in Fig.\,\ref{fig:angcor}\,(c).

A previous $(d,p\gamma)$ experiment reported a multipole-mixing ratio of $\delta \approx 0$ for the primary 2417-keV, $9/2^+ \rightarrow 7/2^-_1$ transition from the 3707-keV state\,\cite{Car70a}, indicating dominant $E1$ character. We confirm the dominant $E1$ character of this transition and determined a multipole-mixing ratio of $\delta = 0.07^{+0.05}_{-0.04}$.

We add two more proton-$\gamma$ angular correlations for the excited states at 3617\,keV [Fig.\,\ref{fig:angcor}\,(i)] and 4661\,keV [Fig.\,\ref{fig:angcor}\,(j)]. The 3617-keV state has an adopted spin-parity assignment of $J^{\pi} = 1/2^-$\,\cite{Jun09a}. Within uncertainties, the measured angular correlation is isotropic as expected for neutron transfer to the $2p_{1/2}$ orbital. See also the discussion above for the 564-keV state. Our data, thus, support the adopted assignment. As discussed, the $\gamma$ decay of the 4661-keV is observed for the first time. The 4661-keV ground-state transition follows the angular correlation expected for a $5/2^- \rightarrow 3/2^-_1$ transition more closely than the one predicted for a $7/2^- \rightarrow 3/2^-_1$ transition [see Fig.\,\ref{fig:angcor}\,(j)]. In fact, $\chi^2_{1f_{7/2}} > \chi^2_{1f_{5/2}}+1$. As the state is at also observed at higher excitation energy, $J^{\pi} = 5/2^-$ is the more likely spin-parity assignment.

\subsection{Reanalysis of $\nuc{52}{Cr}(d,p)\nuc{53}{Cr}$ singles data}
\label{sec:singles}

\begin{figure*}[t]
    \centering
    \includegraphics[width=0.99\linewidth]{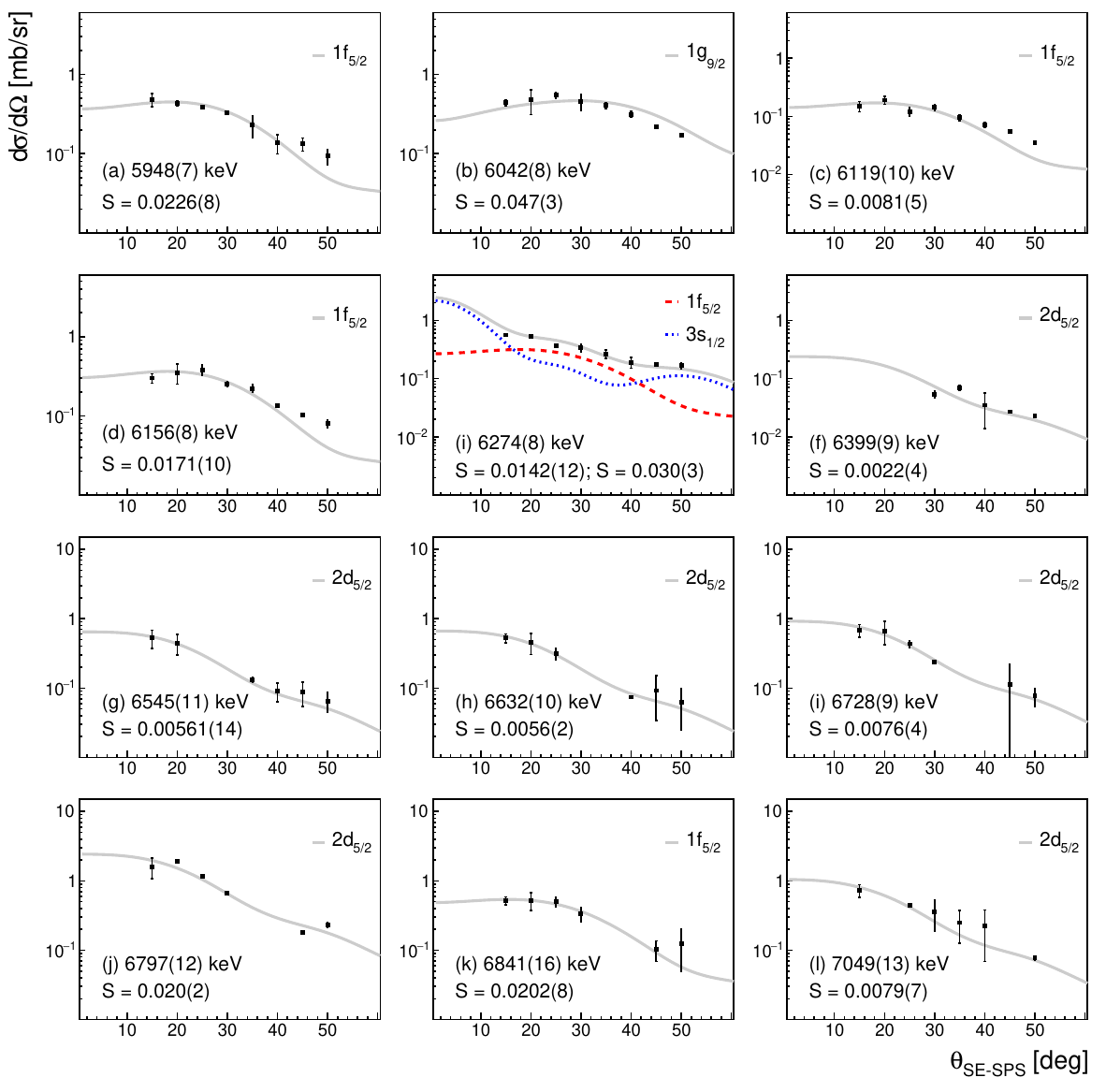}
    \caption{See caption of Fig.\,\ref{fig:angdist_01}.}
    \label{fig:angdist_03}
\end{figure*}

\begin{figure*}[t]
    \centering
    \includegraphics[width=0.99\linewidth]{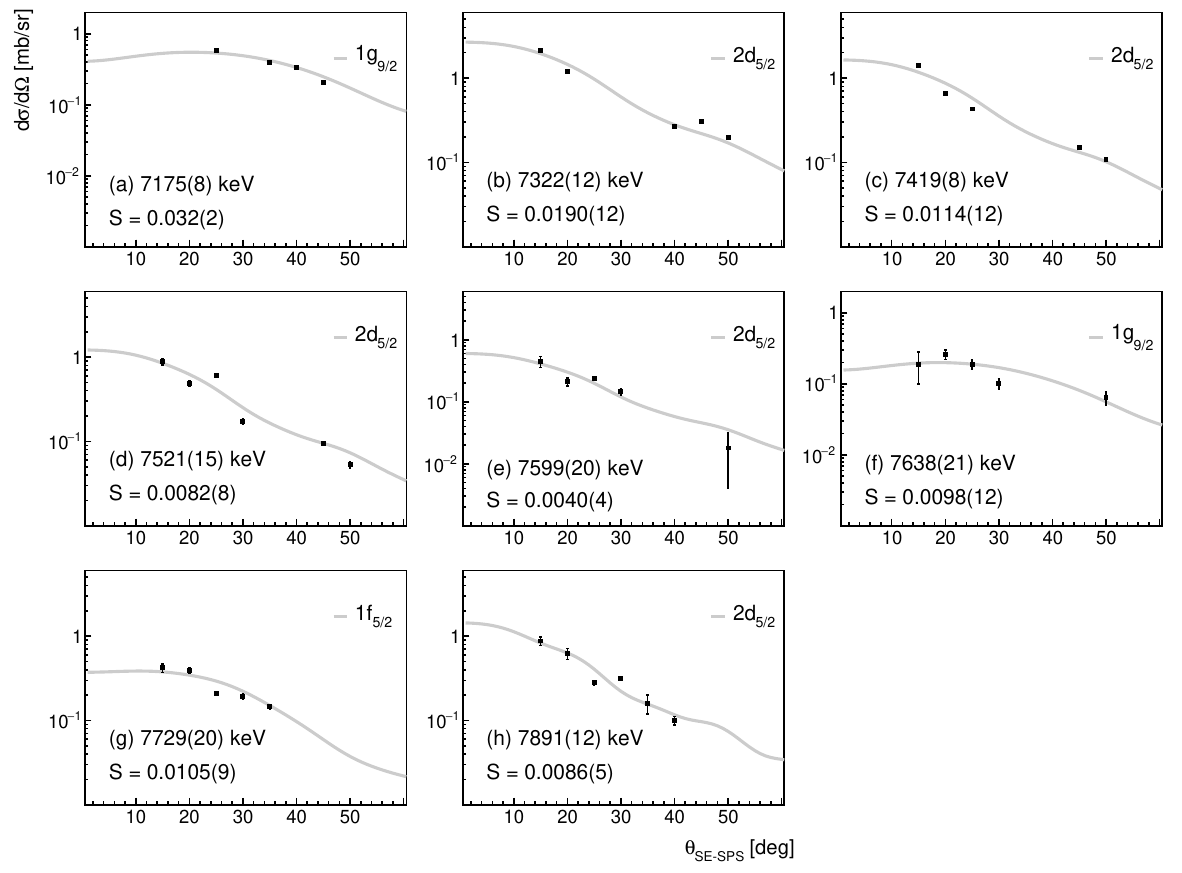}
    \caption{See caption of Fig.\,\ref{fig:angdist_01}.}
    \label{fig:angdist_04}
\end{figure*}

The complementary $\gamma$-ray information and identification of several excited states of \nuc{53}{Cr}, which were not analyzed in \cite{Ril23a}, triggered a reanalysis of the $\nuc{52}{Cr}(d,p)\nuc{53}{Cr}$ angular distributions. In total, 49 excited states were identified up to the neutron-separation energy of $S_n = 7939.43(10)$\,keV\,\cite{Wan21a} and their $(d,p)$ angular distributions measured. The revised angular distributions with corresponding ADWA calculations are shown in Figs.\,\ref{fig:angdist_01} to \ref{fig:angdist_04}. To model these and as also done in \cite{Ril21a, Ril22a, Ril23a, Hay24a}, we assumed that neutrons were transferred into the $2p_{3/2}$, $2p_{1/2}$, $1f_{5/2}$, $1g_{9/2}$, and $2d_{5/2}$ single-particle orbitals above the $N=28$ shell closure. The excitation energies with uncertainties determined in this work are given. The stated uncertainties result from the variation of the excitation energy at the different scattering angles and from the uncertainty of the calibration itself. A second-order polynomial was used for the excitation-energy calibration of the focal plane. In addition, the spectroscopic factors $S =(d\sigma/d\Omega)_{exp}/(d\sigma/d\Omega)_{ADWA}$ including their uncertainties (1$\sigma$-confidence interval) were added to Figs.\,\ref{fig:angdist_01} through \ref{fig:angdist_04}. Data are summarized in table\,\ref{tab:02}. Note that $(d,p)$ with an unpolarized deuteron beam can only determine the angular-momentum transfer, $\ell$. This means that the spin-parity assignment could either be $J^{\pi} = 1/2^-$ or $3/2^-$ for $\ell = 1$, and $J^{\pi} = 5/2^-$ or $7/2^-$ for $\ell=3$ transfers. We chose one of the options for calculating the spectroscopic factor, see Figs.\,\ref{fig:angdist_01}--\ref{fig:angdist_04} and table\,\ref{tab:02}. The choice was guided by either an already available $J^{\pi}$ assignment, complementary $\gamma$-ray data, or systematics available for the other $N=29$ isotones, for some of which experiments with polarized deuterons were performed earlier. Note that, in general, only little $1f_{7/2}$ neutron strength would be expected in the $N=29$ isotones. It is, thus, a valid assumption that most of the $\ell =3$ strengths result from transferring neutrons into the $1f_{5/2}$ orbital.

A few states require a more focused discussion. As already indicated in Fig.\,\ref{fig:matrix}, the 1290-keV state of \nuc{53}{Cr} overlaps with the 3074-keV state of \nuc{54}{Cr}. The latter is populated through an $\ell = 1$ angular momentum transfer. This was now taken into account in the fit. The spectroscopic factor $S$ for the 3074-keV, $J^{\pi} = 2^+$ state, after correcting for the abundance of \nuc{53}{Cr} in natural chromium, is 0.11(3). This is the third $2^+$ state in \nuc{54}{Cr} and all were observed in previous $(d,p)$ experiments\,\cite{ensdf}. The magnitude of the spectroscopic factor appears consistent with the ones determined for $\ell =1$ transfers for the lowest fragments in $\nuc{52}{Cr}(d,p)\nuc{53}{Cr}$, i.e., in our experiment. See Fig.\,\ref{fig:angdist_01}\,(d) for the angular distribution.

The excited states adopted at $E_x = 1537$\,keV and 1974\,keV are located in a region of our $(d,p)$ spectrum [see Fig.\,\ref{fig:xavg}\,(a)] that overlaps with many states of \nuc{54}{Cr} populated in $\nuc{53}{Cr}(d,p)\nuc{54}{Cr}$. This is the $(d,p)$ reaction on our main target contaminant. The energies reported in \cite{Ril23a} strongly deviated from the well-established adopted energies. As the states indeed shift a lot in that region of the $(d,p)$ spectrum when studying them at different scattering angles, we refrain from reporting any data on them in this work.

As already mentioned in Sec.\,\ref{sec:coincidences}, several states are populated around 2.7\,MeV. In our singles data, the 2657-keV and 2670-keV states cannot be resolved. Therefore, we decided to fit the experimental angular distribution with a combination of $\ell = 3$ and $\ell = 1$ transfers populating the 2657-keV and 2670-keV state, respectively. A very good fit of the experimentally measured $(d,p)$ angular distribution is obtained, see Fig.\,\ref{fig:angdist_01}\,(f). Additional evidence for the population of both states is provided in Fig.\,\ref{fig:gammas}\,(a), where $\gamma$-ray transitions from both states are observed. No data were reported for the 2725-keV state, see Fig.\,\ref{fig:angdist_01}\,(g), in \cite{Ril23a}. As already mentioned in Sec.\,\ref{sec:coincidences}, this state corresponds to the previously reported 2723-keV state\,\cite{Jun09a, ensdf}. It could either be a $1/2^-$ or $3/2^-$ state.

No information on the 3164-keV state was provided in \cite{Ril23a} either. We were able to confirm that it is an excited state of \nuc{53}{Cr} and populated in $(d,p)$ because of the complementary $p\gamma$-coincidence data, see Fig.\,\ref{fig:proj}\,(a). Its angular distribution follows that of an $\ell = 1$ transfer, see Fig.\,\ref{fig:angdist_01}\,(h). Given that this state is seen the strongest in the 564-keV gate of Fig.\,\ref{fig:proj}, we believe that this state corresponds to the level adopted at an energy of 3180 keV\,\cite{Jun09a, ensdf}. Obviously, the excitation energies do not agree within uncertainties, which is uncommon in this work for well-established states.

The $(d,p)$ angular distribution for the 3349-keV state is also newly added. It closely follows the one of an $\ell =3$ transfer. Its $\gamma$ decay was discussed in Sec.\,\ref{sec:coincidences} and is consistent with the observed angular-momentum transfer. The state very likely corresponds to the level adopted at 3351 keV with $J^{\pi} = 5/2^-, \ 7/2^-$\,\cite{Jun09a, ensdf}.

As mentioned earlier, excitation energies started deviating significantly from adopted values at around 4.1\,MeV in our previous publication\,\cite{Ril23a}. A careful reanalysis guided by the $p\gamma$-coincidence data solved this problem. For the states adopted at 4135\,keV and 4231\,keV, our excitation energies now agree within uncertainties with the values reported in other $(d,p)$ experiments\,\cite{Rao68a, Boc65a}. We determined an excitation energy of 4129(5)\,keV for the state adopted at 4135.1(6)\,keV. As discussed in Sec.\,\ref{sec:coincidences}, we determined a lower excitation energy of 4220(6)\,keV for the second state, which is consistent with the value of 4223(41)\,keV determined from the $\gamma$-ray  data. As established in previous $(d,p)$ studies, both states are populated through $\ell = 2$ angular-momentum transfers.

Next, we will discuss the group between 4610\,keV and 4661\,keV. While at first glance, this group appears to be a doublet of the 4610-keV and 4661-keV states, based on the angular distributions it turns out to be a triplet of states including the previously observed 4642-keV state as well. Its contribution could not be disentangled by just fitting a 4610-4661 doublet since it contributed to the tails of both peaks. Therefore, the angular distribution of Fig.\,\ref{fig:angdist_02}\,(c), obtained for the entire group, was fitted using three contributions populating the 4610-keV state via an $\ell = 1$, the 4642-keV state via an $\ell =2$, and the 4661-keV state via an $\ell =3$ transfer. Very good agreement was obtained. Our spectroscopic factors qualitatively agree with the ones reported in \cite{Rao68a} as ``Set 2''. Rao {\it et al.} stated $(2J+1)S$ factors of 0.18 for the $\ell =1$, 0.17 for the $\ell = 2$, and 0.68 for the $\ell = 3$ transfers without providing uncertainties. We determined 0.11(2), 0.10(4), and 0.60(12). As discussed in Sec.\,\ref{sec:coincidences}, the observed $p\gamma$-angular correlation supports a $J^{\pi} = 5/2^-$ assignment for the 4661-keV state. From the $\gamma$-ray data, we determined $E_x = 4665(10)$\,keV in agreement with the adopted energy \cite{ensdf}.

\afterpage{
\begingroup
\setlength{\LTcapwidth}{\textwidth} 
\begin{longtable*}[!t]
{
    >{\centering\arraybackslash}p{0.09\textwidth}
    >{\centering\arraybackslash}p{0.09\textwidth}
    >{\centering\arraybackslash}p{0.09\textwidth}
    >{\centering\arraybackslash}p{0.11\textwidth}
    >{\centering\arraybackslash}p{0.11\textwidth}
    >{\centering\arraybackslash}p{0.11\textwidth}
    >{\centering\arraybackslash}p{0.06\textwidth}
    >{\centering\arraybackslash}p{0.06\textwidth}
    >{\centering\arraybackslash}p{0.09\textwidth}
    >{\centering\arraybackslash}p{0.09\textwidth}
}
\caption{\label{tab:02} Summary table for the 
$\nuc{52}{Cr}(d,p)\nuc{53}{Cr}$ singles experiment. Excitation energies, $E_x$, spin-parity assignments, $J^{\pi}$, observed angular momentum transfers, $\ell$, the total spin, $j$, which was used for the ADWA calculations, and spectroscopic factors, $S =(d\sigma/d\Omega)_{exp}/(d\sigma/d\Omega)_{ADWA}$, including their uncertainties (1$\sigma$-confidence interval), are given. For the reanalyzed $\nuc{52}{Cr}(d,p)\nuc{53}{Cr}$ data, results are compared to those of \cite{Ril23a}. As the energy calibration of \cite{Ril23a} is unreliable for $E_x > 4.2$\,MeV, we refrain from comparing the data for excited states of \nuc{53}{Cr} with excitation energies in excess of 4.2\,MeV. The new data supersede the results presented in Ref. \cite{Ril23a}. In the table, ``$-$'' indicates that no information on that observable is available for the state in the associated reference/work. If the field is left empty, we refrain from specifically identifying a previously observed state with a state observed in our work. For the 4610-keV, 4642-keV, and 4661-keV states, the curly bracket in the table indicates that the corresponding angular distribution was fitted with a superposition of three different $\ell$ transfers [see Fig.\,\ref{fig:angdist_02}\,(c)]. For the 5398-keV, 5448-keV, and 6274-keV ``states'', the curly brackets in the table indicate that the associated angular distributions were fitted using a superposition of two different $\ell$ transfers, respectively [see Figs.\,\ref{fig:angdist_02}\,(h) and (i), and Fig.\,\ref{fig:angdist_03}\,(i)]. For the 6274-keV state, the energies of the states within the ``unresolved'' doublet could be determined from our data, see the table and text. Note that the 2657-keV and 2670-keV states were also fitted as an unresolved doublet [see Fig.\,\ref{fig:angdist_01}\,(f)]. That is why no separate excitation energies are reported from the $(d,p)$ singles experiment for them. They could be determined from the observed $\gamma$ decays of these states. See table\,\ref{tab:01}. Additional information can be found in the text and Figs.\,\ref{fig:angdist_01}--\ref{fig:angdist_04}.} \\

\hline
\hline
\multicolumn{3}{c}{$E_x$ [keV]} &
\multicolumn{3}{c}{$J^{\pi}_i$} &
\multicolumn{2}{c}{$\ell,j$} &
\multicolumn{2}{c}{$S$} \\
\cmidrule(lr){1-3}
\cmidrule(lr){4-6}
\cmidrule(lr){7-8}
\cmidrule(lr){9-10}
\multicolumn{1}{c}{\cite{ensdf}} &
\multicolumn{1}{c}{\cite{Ril23a}} &
\multicolumn{1}{c}{This work} &
\multicolumn{1}{c}{\cite{ensdf}} &
\multicolumn{1}{c}{\cite{Ril23a}} &
\multicolumn{1}{c}{This work} &
\multicolumn{1}{c}{\cite{Ril23a}} &
\multicolumn{1}{c}{This work} &
\multicolumn{1}{c}{\cite{Ril23a}} &
\multicolumn{1}{c}{This work} \\
\midrule
\endfirsthead

\multicolumn{10}{c}{\tablename\ \thetable\ -- continued from previous page} \\
\toprule
\multicolumn{3}{c}{$E_x$ [keV]} &
\multicolumn{3}{c}{$J^{\pi}_i$} &
\multicolumn{2}{c}{$\ell,j$} &
\multicolumn{2}{c}{$S$} \\
\cmidrule(lr){1-3}
\cmidrule(lr){4-6}
\cmidrule(lr){7-8}
\cmidrule(lr){9-10}
\multicolumn{1}{c}{\cite{ensdf}} &
\multicolumn{1}{c}{\cite{Ril23a}} &
\multicolumn{1}{c}{This work} &
\multicolumn{1}{c}{\cite{ensdf}} &
\multicolumn{1}{c}{\cite{Ril23a}} &
\multicolumn{1}{c}{This work} &
\multicolumn{1}{c}{\cite{Ril23a}} &
\multicolumn{1}{c}{This work} &
\multicolumn{1}{c}{\cite{Ril23a}} &
\multicolumn{1}{c}{This work} \\
\midrule
\endhead

\midrule
\multicolumn{10}{r}{{Continued on next page}} \\
\endfoot

%\bottomrule
\hline
\hline
\endlastfoot

0    & 0      & 0      & $3/2^-$ & $3/2^-$ & $3/2^-$ & $1,3/2$ & $1,3/2$ & 0.33(2) & 0.26(2) \\
564.03(4) & 564(2) & 565(2) & $1/2^-$ & $1/2^-$ & $1/2^-$ & 1,1/2 & 1,1/2 & 0.21(2) & 0.185(11) \\
1006.27(5) & 1006(2) & 1006(3) & $5/2^-$ & $5/2^-$ & $5/2^-$ & 3,5/2 & 3,5/2 & 0.21(1) & 0.181(5) \\
1289.52(7) & 1289(2) & 1288(4) & $7/2^-$ & $7/2^-$ & $7/2^-$ & 3,7/2 & 3,7/2 & 0.032(3) & 0.033(4) \\
1536.62(7) & 1549(11) & $-$ & $7/2^-$ & $7/2^-$ & $-$ & 3,7/2 & $-$ & 0.008(1) & $-$ \\
1973.66(11) & 1949(12) & $-$ & $5/2^-$ & $1/2^-,3/2^-$ & $-$ &1,1/2 & $-$ & 0.11(2) & $-$ \\
2320.7(2) & 2317(4) & 2319(4) & $3/2^-$ & $3/2^-$ & $3/2^-$ & $1,3/2$ & $1,3/2$ & 0.15(1) & 0.1573(9) \\
2656.5(3) & 2664(6) & $-$ & $5/2^-,7/2^-$ & $5/2^-,7/2^-$ & $5/2^-,7/2^-$ & $3,5/2$ & $3,5/2$ & 0.11(1) & 0.088(8) \\
2669.9(5) & $-$ & $-$ & $1/2^-$ & $-$ & $1/2^-$ & $-$ & $1,1/2$ & $-$ & 0.017(3) \\
2723(10) & $-$ & 2725(7) & $1/2^-,3/2^-$ & $-$ & $1/2^-,3/2^-$ & $-$ & $1,1/2$ & $-$ & 0.0174(2) \\ 
 3180.05(21) & $-$ & 3164(7) & $(3/2)^-$ & $-$ & $3/2^-$ & $-$ & $1,3/2$ & $-$ & 0.0094(5) \\
 3351(6) & $-$ & 3349(5) & $5/2^-,7/2^-$ & $-$ & $5/2^-,7/2^-$ & $-$ & $3,5/2$ & $-$ & 0.0068(8) \\
 3616.51(18) & 3619(9) & 3618(2) & $1/2^-$ & $1/2^-$ & $1/2^-$ & $1,1/2$ & $1,1/2$ & 0.20(2) & 0.16(2) \\
 3706.5(15) & 3712(10) & 3709(3) & $9/2^+$ & $9/2^+$ & $9/2^+$ & $4,9/2$ & $4,9/2$ & 0.22(1) & 0.234(14) \\
 4135.1(6) & 4170(11) & 4129(5) & $5/2^+,3/2^+$ & $5/2^+$ & $5/2^+$ & $2,5/2$ & $2,5/2$ & 0.054(4) & 0.036(2) \\
4230.5(7) & 4268(10) & 4220(6) & $5/2^+,3/2^+$ & $5/2^+$ & $5/2^+$ & $2,5/2$ & $2,5/2$ & 0.027(2) & 0.0176(14) \\
 &  & 4495(5) &  & & $5/2^+$ & & $2,5/2$ &  & 0.0085(6) \\ 
4610(7) & & $-$ & $1/2^-,3/2^-$ & & $1/2^-,3/2^-$ & \multirow{3}{*}{$\Bigg\{$}  & $1,1/2$ & & 0.055(11) \\
4642(7) & & $-$ & $5/2^+,3/2^+$ & & $5/2^+$ & & $2,5/2$ & & 0.016(7) \\
4661(7) & & $-$ &$5/2^-,7/2^-$ & & $5/2^-$ & & $3,5/2$ & & 0.10(2) \\
 & & 4803(5) &  & & $5/2^+$ & & $2,5/2$ & & 0.0044(5) \\
5123(10) & & 5122(5) & $3/2^+,5/2^+$ & & $5/2^+$ & & $2,5/2$ & & 0.00236(11) \\
 & & 5180(6) & & & $5/2^-,7/2^-$ & & $3,5/2$ & & 0.0187(5) \\
 & & 5259(6) & & & $5/2^-,7/2^-$ & & $3,5/2$ & & 0.016(2) \\
 5397(10) & & 5398(8) & $1/2^-,3/2^-$ & & $1/2^-,3/2^-$ & \multirow{2}{*}{\hfill$\Big\{$} & $1,1/2$ & & 0.013(2) \\
 5420(10) & & $-$ & $3/2^+,5/2^+$ & & $5/2^+$ & & $2,5/2$ & & 0.0056(3) \\
 5452(10) & & 5448(8) & $1/2^-,3/2^-$ & & $1/2^-,3/2^-$ & \multirow{2}{*}{\hfill$\Big\{$} & $1,1/2$ & & 0.010(2) \\
  & & $-$ & & & $5/2^+$ & & $2,5/2$ & & 0.0016(3) \\
  & & 5533(6) & & & $5/2^+$ & & $2,5/2$ & & 0.00111(10) \\
  & & 5577(5) & & & $5/2^+$ & & $2,5/2$ & & 0.0040(3) \\
  & & 5857(11)& & & $5/2^-,7/2^-$ & & $3,5/2$ & & 0.0065(4) \\
  & & 5948(7) & & & $5/2^-,7/2^-$ & & $3,5/2$ & & 0.0226(8) \\
  & & 6042(8) & & & $9/2^+$ & & $4,9/2$ & & 0.047(3) \\
  & & 6119(10) & & & $5/2^-,7/2^-$ & & $3,5/2$ & & 0.0081(5) \\
  & & 6156(8) &  & & $5/2^-,7/2^-$ & & $3,5/2$ & & 0.0171(10) \\
  & & 6252(11)&  & & $5/2^-,7/2^-$ & \multirow{2}{*}{$\Big\{$} & $3,5/2$ & & 0.030(3) \\
  6305(10)& & 6297(11)& $1/2^+$ & & $1/2^+$ & & $1,1/2$& & 0.0142(12) \\
  & & 6399(9) &  & & $5/2^+$ & & $2,5/2$ & & 0.0022(4) \\
  6524(10)& & 6545(11)& $(3/2,5/2)^+$ & & $5/2^+$ & & $2,5/2$ & & 0.00561(14) \\
  6630(10) & & 6632(10)& $(3/2,5/2)^+$& & $5/2^+$ & & $2,5/2$ & & 0.0056(20) \\
  6735(10) & &6728(9)& $(3/2,5/2)^+$ & & $5/2^+$ & & $2,5/2$ & & 0.0076(4) \\
  & & 6797(12) & & & $5/2^+$ & & $2,5/2$ & & 0.020(2) \\
  & & 6841(16) & & & $5/2^-,7/2^-$ & & $3,5/2$ & & 0.0202(8) \\
  & & 7049(13) & & & $5/2^+$ & & $2,5/2$ & & 0.0079(7) \\
  & & 7175(8)  & & & $9/2^+$ & & $4,9/2$ & & 0.032(2) \\
  & & 7322(12) & & & $5/2^+$ & & $2,5/2$ & & 0.0190(12) \\
  & & 7419(8)  & & & $5/2^+$ & & $2,5/2$ & & 0.0114(12) \\
  & & 7521(15) & & & $5/2^+$ & & $2,5/2$ & & 0.0082(8) \\
  & & 7599(20) & & & $5/2^+$ & & $2,5/2$ & & 0.0040(4) \\
  & & 7638(21) & & & $9/2^+$ & & $4,9/2$ & & 0.0098(12) \\
  & & 7729(20) & & & $5/2^-,7/2^-$ & & $3,5/2$ & & 0.0105(9) \\
  & & 7891(12) & & & $5/2^+$ & & $2, 5/2$ & & 0.0086(5)
\end{longtable*}
\endgroup
}

Comments are necessary for the excited states at 5398\,keV and 5448\,keV, see Figs.\,\ref{fig:angdist_02}\,(h) and (i). As did Bock {\it et al.}\,\cite{Boc65a}, we do not observe an angular distribution for the 5398-keV state which can be explained by one angular momentum transfer. While they decided to fit one angular-momentum transfer ($\ell =1$) to their measured angular distribution, we chose to fit a superposition of $\ell = 1$ and $\ell = 2$ transfers to this ``state'', which would indicate the presence of a doublet in our case. Bock {\it et al.} indeed reported two close-lying states at 5388\,keV and 5412\,keV, which in their study were populated through $\ell = 1$ and $\ell =2$ transfers and which they were able to fit separately. We cannot reliably resolve these two close-lying states. The average energy would be 5400\,keV and, thus, the angular distribution observed in our work results likely from the superposition of the two states with the excitation energy being consistent. Bock {\it et al.} also reported a state at 5448\,keV and fitted its angular distribution with an $\ell = 1$ transfer. A state at that energy is also observed in our work. Its angular distribution is neither fitted well by an $\ell = 1$ nor an $\ell = 2$ transfer. The $\ell = 2$ transfer would, however, provide the slightly better fit in our case. In this case, the spectroscopic factor would be $S = 0.0032(3)$ for the neutron transfer to the $2d_{5/2}$ orbital. The fit with the superposition of the two, see Fig.\,\ref{fig:angdist_02}\,(i), is not satisfactory. It results, however, in a better fit than any single $\ell$ transfer and would indicate the presence of a doublet that was neither resolved in our work nor in previous work. The spectroscopic factors for the fit with both $\ell$ transfers are reported in Fig.\,\ref{fig:angdist_02}\,(i). For this group of states, there is a possibility that we observe the combined effects coming from populating the adopted states at 5397(10)\,keV, 5420(10)\,keV, and 5452(10)\,keV \cite{ensdf}. See also table\,\ref{tab:02}. The adopted energies do, however, not correspond to the ones reported by Bock {\it et al.} \cite{Boc65a}.

Bock {\it et al.} reported the population of several $J^{\pi} = 1/2^+$ states via $\ell = 0$ transfers in their $(d,p)$ experiment\,\cite{Boc65a} with the first one for a state at 4.684\,MeV. Rao {\it et al.} observed the state at 4.696\,MeV and also proposed that it was populated via an $\ell = 0$ transfer\,\cite{Rao68a}, even though their fit appears less convincing than the one of \cite{Boc65a}. This state would appear in the tail of the 4610-4661 group discussed earlier. Adding an $\ell =0$ transfer was not necessary to explain our data, see Fig.\,\ref{fig:angdist_02}\,(c). Rao {\it et al.} also reported tentative $J^{\pi} = 1/2^+$ assignments for states which they observed at 4.435\,MeV and 4.489\,MeV. We do not observe a candidate for the first but we do see a candidate for the second state at 4.495(5)\,MeV. Our angular distribution is fitted well assuming an $\ell =2$ transfer, see Fig.\,\ref{fig:angdist_02}\,(b). For higher-lying states, Bock {\it et al.} report the next $\ell = 0$ transfer at 6.282\,MeV\,\cite{Boc65a}. We observe a state at 6.274(8)\,MeV. The experimental angular distribution is fitted better when including an $\ell = 0$ in addition to an $\ell = 3$ angular-momentum transfer. Again, this would indicate the existence of a doublet which could not be clearly resolved in our experiment. It is possible that the other component is the 6.247-MeV state, which Bock {\it et al.} reported\,\cite{Boc65a}. If we fit a doublet, we establish states at 6.252(11)\,MeV and 6.297(11)\,MeV, which agree well with the excitation energies reported in \cite{Boc65a}. The 6252-keV state would be populated through an $\ell =3$ and the 6297-keV state through an $\ell = 0$ transfer. Bock {\it et al.} also reported $\ell = 0$ transfers for states at 6.659\,MeV, 6.688\,MeV, 6.832\,MeV, 6.960\,MeV, and 7.171\,MeV. We do not observe a candidate for the 6.960-MeV state with the caveat that it would be masked by states of \nuc{13}{C} at many angles. The population of states which we observe at 6.632(10)\,MeV, 6.841(16)\,MeV, and 7.175(8)\,MeV is well described assuming $\ell=2$, $\ell=3$, and $\ell =4$ transfers, respectively. It should be noted though that two of the states likely correspond to the 6.626-MeV state, for which $\ell =2$ was listed, and the 6.853-MeV state reported in \cite{Boc65a} rather than the $J^{\pi} = 1/2^+$ candidates. In conclusion, our work can only support the existence of a $J^{\pi} = 1/2^+$ state at 6297(11)\,keV. We also note that this is consistent with not having observed any significant $\ell =0$ strengths in our previous $(d,p)$ work in this mass region\,\cite{Ril21a, Ril22a, Ril23a, Spi23a}.

\begin{figure}[t]
    \centering
    \includegraphics[width=0.99\linewidth]{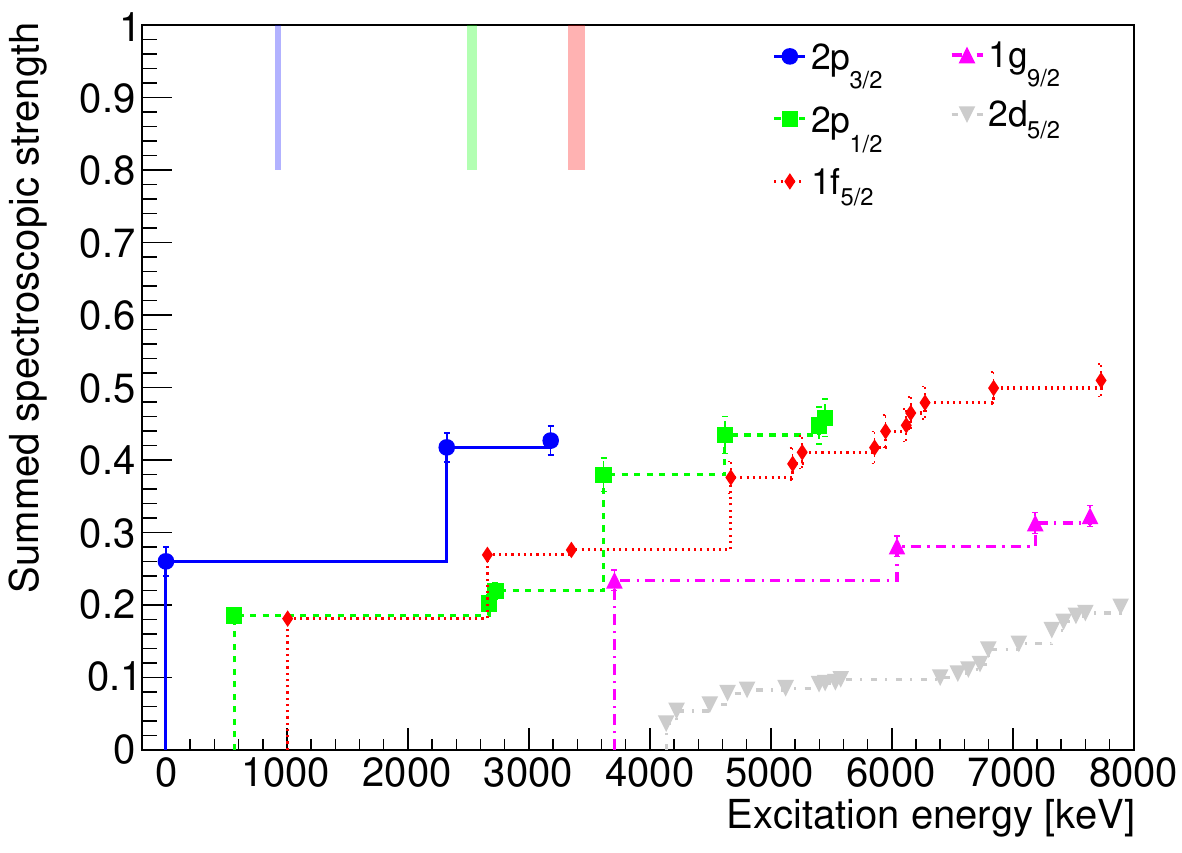}
    \caption{Summed spectroscopic strengths measured in $\nuc{52}{Cr}(d,p)\nuc{53}{Cr}$ for placing neutrons in the $2p_{3/2}$, $2p_{1/2}$, $1f_{5/2}$, $1g_{9/2}$, and $2d_{5/2}$ single-particle orbitals. The vertical lines indicate the effective single-particle energy for the respective orbital. The color coding is the same. The width of the line corresponds to the uncertainty of the centroid energy. No centroids were calculated for the $1g_{9/2}$ and $2d_{5/2}$ neutron orbitals.}
    \label{fig:runsum}
\end{figure}

Most states observed above an excitation energy of 4\,MeV were populated in $(d,p)$ via $\ell = 2$ transfers. The running sums for the spectroscopic strengths up to the neutron-separation energy and attributed to the filling of the neutron $2p_{3/2}$, $2p_{1/2}$, $1f_{5/2}$, $1g_{9/2}$, and $2d_{5/2}$ single-particle orbitals are shown in Fig.\,\ref{fig:runsum}. While we collect 43(2)\,$\%$ of the sum-rule strength for the $2p_{3/2}$, 46(5)\,$\%$ for the $2p_{1/2}$, and 51(4)\,$\%$ for the $1f_{5/2}$ orbitals, only 32(2)\,$\%$ and 20(2)\,$\%$ of the spectroscopic strengths are observed to be concentrated in resolved states for the $1g_{9/2}$ and $2d_{5/2}$ orbitals, respectively. Note that within uncertainties the value for the $1g_{9/2}$ orbital is consistent with the one reported in \cite{Ril23a, Hay24a} even though two additional, smaller fragments were added up to $S_n$. Obviously, while the spectroscopic strengths for the $p$ orbitals and $1f_{5/2}$ orbital are consistent with the expected quenching of the single-particle strength\,\cite{Kay13a, Kay22a}, significantly less strength is collected for the $1g_{9/2}$ and $2d_{5/2}$ orbitals. In contrast to all other single-particle orbitals, there is no ``strong'' fragment for the $2d_{5/2}$. We also see that the $2d_{5/2}$ strength is fragmented among a significantly larger number of states. It is instructive to mention that 28$\%$ of the observed $1f_{5/2}$ strength is collected by smaller fragments. Thus, while there are stronger fragments, a significant part of the $1f_{5/2}$ strength is carried by smaller fragments. We hypothesized earlier that continuum physics might play a role for the $1g_{9/2}$ orbital as it is predicted to be unbound by covariant density functional theory (CDFT) calculations employing the relativistic functional FSUGarnet\,\cite{Che15a} or that there might be smaller fragments, which could not be resolved. Comparing to the $1f_{5/2}$ and $2d_{5/2}$ strengths, it can be expected that there are many smaller fragments for the $1g_{9/2}$ strength. It is, however, likely that many of these fragments will be above the neutron-separation energy. Therefore, it cannot be excluded that continuum effects might play a prominent role and could possibly quench the strength even further. Certainly, studies with highly-enriched and clean targets would be needed to study these states as many broad states of carbon and oxygen will be populated above the neutron-separation energy of stable nuclei around the $N=28$ shell closure. Coincidence experiments as presented in this work might help as well, if the states still have a significant $\gamma$-decay width.

The following centroid energies were determined in this work: 926(44)\,keV for the $2p_{3/2}$, 2529(85) for the $2p_{1/2}$, and 3379(59)\,keV for the $1f_{5/2}$ orbital. These effective single-particle energies are shown as vertical lines in Fig.\,\ref{fig:runsum}. The effective single-particle energies lead to a spin-orbit splitting of $\Delta SO = E_{2p_{1/2}} - E_{2p_{3/2}} = 1603(96)$\,keV and and a splitting between the pseudospin partners of $\Delta PS = E_{1f_{5/2}} - E_{2p_{3/2}} = 2453(73)$\,keV, respectively. Note that pseudospin symmetry is a feature of relativistic mean-field theories that have an attractive scalar and repulsive vector potential nearly equal in magnitude \cite{Gin97a}. Relative to the neutron-separation energy of \nuc{53}{Cr}, the single-particle energies are 7.01(4)\,MeV for the $2p_{3/2}$, 5.41(9)\,MeV for the $2p_{1/2}$, and 4.56(6)\,MeV for the $1f_{5/2}$. The CDFT calculations with FSUGarnet predict 6.65\,MeV, 5.39\,MeV, and 4.73\,MeV in good agreement with the experimental values. We add that the experimental binding energy for the $2p_{3/2}$, $S_n - E_{2p_{3/2}} = 7.01(4)$\,MeV, and the spin-orbit splitting $\Delta SO = 1.60(10)$\,MeV agree well with the trend established in \cite{Che24a}. FSUGarnet predicts $\Delta SO = 1.26$\,MeV. This difference to the experimental value can be attributed to the $2p_{3/2}$ being less bound in theory than in experiment. For comparison, we add the single-particle energies predicted with the RMF022 and RMF028 relativistic functionals \cite{Che15a} to table\,\ref{tab:03}. We see that all three functionals predict a spin-orbit splitting between 1.25\,MeV and 1.26\,MeV. The predictions for the splitting between the pseudospin partners, $\Delta PS = 1f_{5/2} - 2p_{3/2}$, and for the gap between the $1f_{5/2}$ and $2p_{1/2}$ neutron orbitals, $\Delta_{pf} = E_{1f_{5/2}}-E_{2p_{1/2}}$, are quite different though, see table\,\ref{tab:03}. Overall, the predictions of FSUGarnet appear to agree best with the experimental data for \nuc{53}{Cr}.

\begin{figure}[t]
    \centering
    \includegraphics[width=0.99\linewidth]{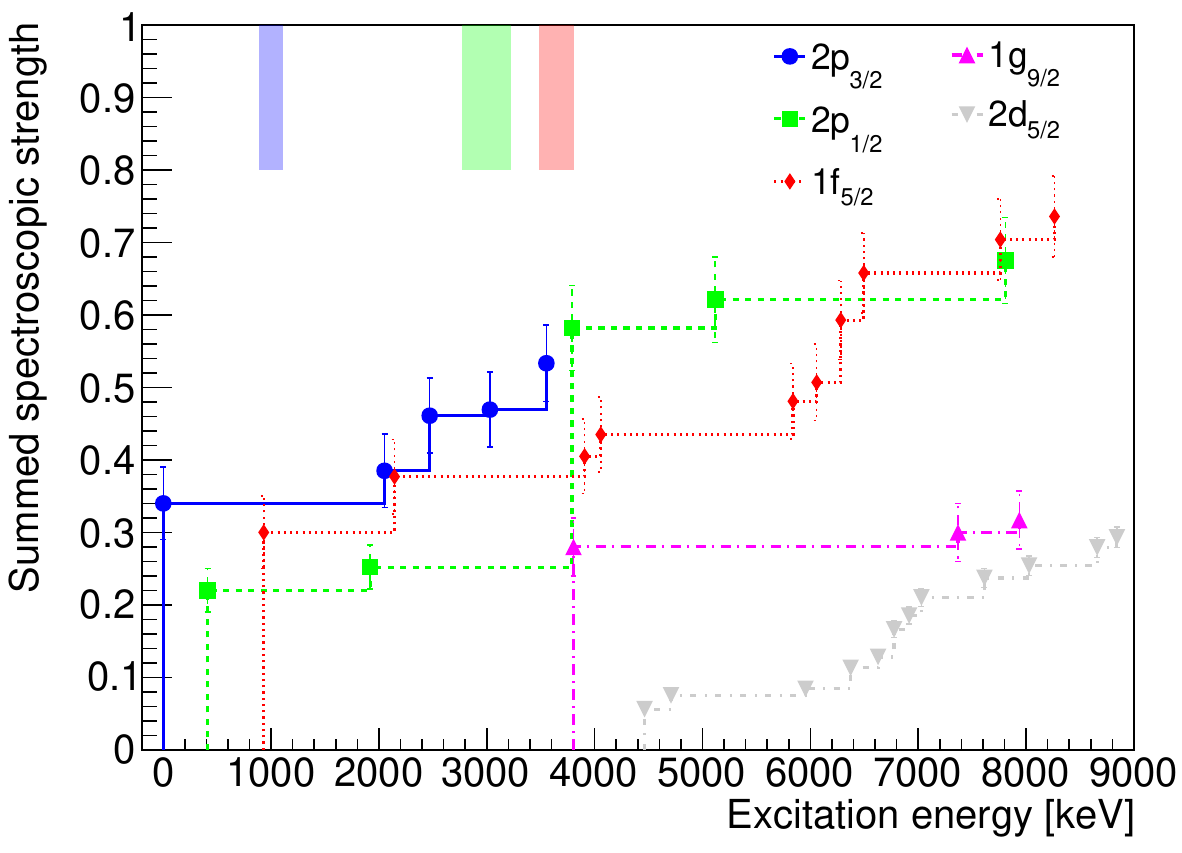}
    \caption{Same as Fig.\,\ref{fig:runsum} but for $\nuc{54}{Fe}(d,p)\nuc{55}{Fe}$\,\cite{Ril22a}.}
    \label{fig:runsum_fe}
\end{figure}

\begingroup
\squeezetable
\renewcommand*{\arraystretch}{1.2}
\begin{table*}[!t]
\caption{\label{tab:03} Comparison of the energies of the $2p_{3/2}$, $2p_{1/2}$, and $1f_{5/2}$ neutron single-particle orbitals. Shown are experimentally determined values and predictions by FSUGarnet, RMF022, and RMF028 \cite{Che15a}. Energies are given relative to $S_n$, $e.g.$, $2p_{3/2} = S_n - E_{2p_{3/2}}$, where $E_{2p_{3/2}}$ would correspond to the centroid energy of the single-particle orbit. The spin-orbit splitting, $\Delta SO = 2p_{1/2} - 2p_{3/2}$, the splitting between the pseudospin partners \cite{Gin97a}, $\Delta PS = 1f_{5/2} - 2p_{3/2}$, and the gap between the $1f_{5/2}$ and $2p_{1/2}$ orbitals, $\Delta_{pf} = E_{1f_{5/2}}-E_{2p_{1/2}}$, are also given. See text for further details.}
\begin{ruledtabular}
\begin{tabular}{ccccccc}

Type & $2p_{3/2}$ & $2p_{1/2}$ & $1f_{5/2}$ & $\Delta SO$ & $\Delta PS$ & $\Delta_{pf}$ \\
 & [MeV] & [MeV] & [MeV] & [MeV] & [MeV] \\
\hline
\multicolumn{7}{c}{\nuc{53}{Cr}} \\
Exp. & 7.01(4) & 5.41(9) & 4.56(6) & 1.60(10) & 2.45(7) & 0.85(10) \\
FSUGarnet & 6.65 & 5.39 & 4.73 & 1.26 & 1.92 & 0.66 \\
RMF022 & 6.75 & 5.47 & 4.94 & 1.26 & 1.78 & 0.52 \\
RMF028 & 6.84 & 5.59 & 5.26 & 1.25 & 1.58 & 0.324\\
\multicolumn{7}{c}{\nuc{55}{Fe}} \\
Exp. & 8.22(12) & 6.3(2) & 5.6(2) & 1.9(2) & 2.6(2) & 0.66(27)\\
FSUGarnet & 7.79 & 6.58 & 6.44 & 1.20 & 1.33 & 0.14\\
RMF022 & 7.79 & 6.58 & 6.61 & 1.21 & 1.17 & -0.03\\
RMF028 & 7.84 & 6.63 & 6.87 & 1.21 & 0.97 & -0.24
 
\end{tabular}
\end{ruledtabular}
\end{table*}
\endgroup

Note that the revised experimental values do not significantly alter the discussions presented in \cite{Ril23a, Hay24a, Spi24a} and the discussed change of the $2p_{1/2}$-$1f_{5/2}$ gap, $\Delta_{pf}$, with changing proton number $Z$. For comparison, we add the running sums of the spectroscopic strengths for $\nuc{54}{Fe}(d,p)\nuc{55}{Fe}$\,\cite{Ril22a}, see Fig.\,\ref{fig:runsum_fe}. In this case, the effective single-particle energies are 1077(116)\,keV for the $2p_{3/2}$, 2998(221)\,keV for the $2p_{1/2}$, and 3653(159)\,keV for the $1f_{5/2}$ orbitals. Based on a comparison to \nuc{53}{Cr}, we decided to consider the 7938-keV state in \nuc{55}{Fe} as a $J^{\pi} = 9/2^+$ state. The gap between the $2p_{1/2}$ and $1f_{5/2}$ is 655(272)\,keV in \nuc{55}{Fe} and 851(104)\,keV in \nuc{53}{Cr}. We added this data to table\,\ref{tab:03}. Based on our data reported in \cite{Ril21a}, $\Delta_{pf} = 1712(216)$\,keV in \nuc{51}{Ti}. Even though there are large uncertainties involved, the gap between the $2p_{1/2}$ and $1f_{5/2}$ neutron orbitals clearly appears to get smaller with increasing proton number $Z$. The CDFT calculations with FSUGarnet predict 1.2\,MeV for \nuc{51}{Ti}\,\cite{Ril21a}, 0.66\,MeV for \nuc{53}{Cr}, and 0.14\,MeV for \nuc{55}{Fe} and, thus, qualitatively support the closing of the gap. The other functionals also predict a closing of the gap with increasing $Z$. However, for RMF022 and RMF028, the $1f_{5/2}$ orbital drops below the $2p_{1/2}$ orbital in \nuc{55}{Fe}, which is not observed experimentally (see table\,\ref{tab:03} and Fig.\,\ref{fig:runsum_fe}). For the $2p$ orbitals in \nuc{55}{Fe}, we get $S_n - E_{2p_{3/2}} = 8.22(12)$\,MeV and the spin-orbit splitting $\Delta SO = 1.92(25)$\,MeV. The latter is also consistent with the trend reported in \cite{Che24a}, i.e., that the spin-orbit splitting should increase with increased binding of the $2p_{3/2}$ orbital. The CDFT calculations with FSUGarnet predict a spin-orbit splitting between the $p$ orbitals of 1.20\,MeV, i.e., expect it to be smaller than in \nuc{53}{Cr}. RMF022 and RMF028 both predict $\Delta SO = 1.21$\,MeV, i.e., a value consistent with the prediction of FSUGarnet. The predictions are, however, different from what is observed in experiment, even though the CDFT calculations with FSUGarnet correctly describe that the $2p_{3/2}$ is more strongly bound in \nuc{55}{Fe} than in \nuc{53}{Cr}\,\cite{Ril23a}.

\begin{figure}[b]
    \centering
    \includegraphics[width=0.99\linewidth]{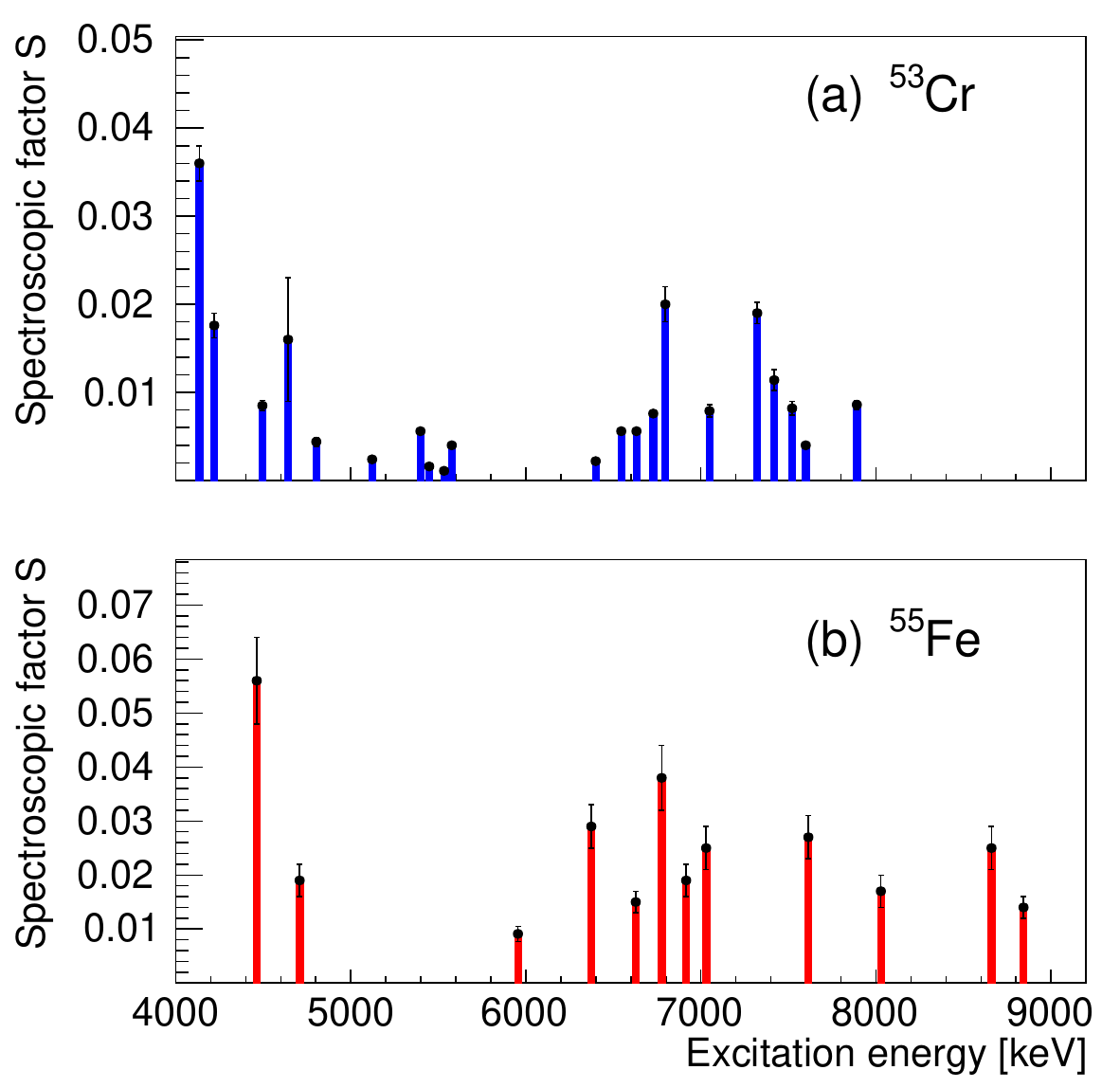}
    \caption{Spectroscopic factors $S$ observed for excited states of (a) \nuc{53}{Cr} and (b) \nuc{55}{Fe} populated via $\ell = 2$ angular momentum transfer in the $(d,p)$ reaction. Neutron transfer to the $2d_{5/2}$ orbital was assumed.}
    \label{fig:d52}
\end{figure}

In closing, we want to mention that at first glance the fragmentation of the $2d_{5/2}$ strength appears equally strong in \nuc{55}{Fe} and \nuc{53}{Cr} (see Figs.\,\ref{fig:runsum} and \ref{fig:runsum_fe}). The details of the strength fragmentation are, however, different (see Fig.\,\ref{fig:d52}). Interestingly, the strength fragmentation for the other orbitals is quite similar in \nuc{53}{Cr} and \nuc{55}{Fe}. While we observe the $2d_{5/2}$ neutron strength to be fragmented among 21 states for \nuc{53}{Cr}, it is only fragmented among 12 states in \nuc{55}{Fe}. Note that the ``two-groups'' structure resembles what we observed earlier for \nuc{62}{Ni}\,\cite{Spi23a}. It is quite possible that the valence protons play a pivotal role in determining how the strength gets fragmented for the neutron $2d_{5/2}$ orbital. In first order, one might expect that the density of states is higher around proton mid-shell, i.e., that it is higher for \nuc{53}{Cr} ($Z=24$) than \nuc{55}{Fe} ($Z=26$). At the energies, where we observe the $J^{\pi} = 5/2^+$ states, $(\pi:1d_{5/2})^{-1}(\pi:1f_{7/2})^{+1}$ one-particle-one-hole (1p-1h) excitations could be energetically possible and mix with the neutron states. One could search for these with proton removal reactions. Another possibility, as already discussed in \cite{Ril23a} for the $1g_{9/2}$ strength, is that we get $J^{\pi} = 5/2^+$ states from the coupling of the collective octupole phonon with the unpaired neutron, i.e., from configurations like $(3^- \otimes 2p_{3/2})_{5/2^+}$. In \nuc{52}{Cr}, the collective octupole excitation is at 4.56\,MeV. In \nuc{54}{Fe}, it is at 4.78\,MeV\,\cite{Kib02a}. Therefore, the phonon energy appears to be in the energy range of the neutron $2d_{5/2}$ single-particle strength. At the same time, one has to be careful as very likely the $(\nu:1f_{7/2})^{-1}(\nu:2d_{5/2})^{+1}$ and $(\nu:2p_{3/2})^{-1}(\nu:2d_{5/2})^{+1}$ 1p-1h excitations contribute to the structure of the phonon itself. Without realistic theoretical calculations, one cannot judge whether mixing with complex configurations and effects of Pauli blocking will play a role in quenching the spectroscopic strengths at higher energies. Calculations along these lines would be instructive.

\section{Summary}

A $\nuc{52}{Cr}(d,p\gamma)\nuc{53}{Cr}$ experiment was performed with the combined SE-SPS and CeBrA demonstrator setup at the FSU John D. Fox Accelerator Laboratory. The complementary information triggered a reanalysis of our previous $(d,p)$ work\,\cite{Ril23a}. As a result, 49 excited states of \nuc{53}{Cr} were firmly identified to be populated in the $\nuc{52}{Cr}(d,p)\nuc{53}{Cr}$ reaction even though a natural chromium target was used. For all states, $(d,p)$ angular distributions were measured and through comparison to ADWA calculations spectroscopic factors extracted. The latter were used to reassess the fragmentation of the $2p_{3/2}$, $2p_{1/2}$, $1f_{5/2}$, $1g_{9/2}$, and $2d_{5/2}$ single-particle strengths in \nuc{53}{Cr}. Up to the neutron-separation energy of \nuc{53}{Cr}, we collected 43(2)\,$\%$ of the sum-rule strength for the $2p_{3/2}$, 46(5)\,$\%$ for the $2p_{1/2}$, and 51(4)\,$\%$ for the $1f_{5/2}$ orbitals. The decrease of the $2p_{1/2}-1f_{5/2}$ shell gap with increasing proton number $Z$ was confirmed in the $N=29$ isotones. For the $1g_{9/2}$ and $2d_{5/2}$ single-particle orbitals, only 32(2)\,$\%$ and 20(2)\,$\%$ of the spectroscopic strengths were observed to be concentrated in resolved states, respectively. These numbers are consistent with results for \nuc{55}{Fe} and a comparison was presented in this work. Studies with highly-enriched and clean targets without light backings would be needed to find even weaker fragments of the $1g_{9/2}$ and $2d_{5/2}$ neutron single-particle strengths since many broad states of carbon and oxygen will be populated above the neutron-separation energy of stable nuclei around the $N=28$ shell closure. Coincidence experiments as presented in this work might help if the states above the threshold still have a significant $\gamma$-decay width. Possibly, $\gamma$-ray gates similar to the ones employed in this work but gating on $\gamma$ rays of the residual nucleus after neutron emission of states above $S_n$ could be used, too. We briefly commented that mixing with more complex configurations might drive the fragmentation of the neutron single-particle strengths at higher energies. We also hypothesized, using the data for the neutron $2d_{5/2}$ strength in \nuc{53}{Cr} and \nuc{55}{Fe}, that mixing and possibly connected Pauli blocking could lead to further quenching of the spectroscopic strength. It is, however, important to point out that no multi-step contributions were needed to describe the $(d,p)$ angular distributions measured in our work. All distributions could be described assuming direct transfer of the neutron. Additional studies above $N=28$ seem instructive to provide the data for constructing a shell-model cross-shell interaction that also includes the $g$ and $d$ orbitals, i.e., similar in spirit to the FSU interaction\,\cite{Lub20a}. The development of such an interaction is crucial for spectroscopy studies hoping to connect the $N=28$ and $N=40$ islands of inversion\,\cite{Bro25a} and for, {\it e.g.}, understanding recent data in the Ge-Se mass region\,\cite{Spi24b}. Experimental studies should also include nucleon removal reactions to understand both the occupancies of orbitals and role of complex configurations.

\begin{acknowledgments}
This work was supported by the U.S. National Science Foundation (NSF) under Grant Nos. PHY-2012522 (FSU) [WoU-MMA: Studies of Nuclear Structure and Nuclear Astrophysics], PHY-2405485 [MRI: Equipment: A New CeBr$_3$ Gamma-Ray Detection Array (CeBrA) for Particle-Gamma Coincidence Experiments at the FSU Super-Enge Split-Pole Spectrograph], and PHY-2412808 (FSU) [Studies of Nuclear Structure and Nuclear Astrophysics]. The work was also supported by the U.S. Department of Energy, Office of Science, Office of Nuclear Physics under under Award No. DE-FG02-92ER40750 (FSU). Support by Florida State University is also gratefully acknowledged.
\end{acknowledgments}

\bibliography{references}% Produces the bibliography via BibTeX.

\end{document}